\documentclass[journal]{IEEEtran}
\usepackage{cite}
\usepackage{times,amsmath,amssymb,psfrag}
\usepackage{graphicx}
\usepackage{url}
\usepackage{bm}
\usepackage{multicol}
\usepackage{upgreek}
\allowdisplaybreaks[4]
\graphicspath{{./pics/}} \DeclareGraphicsExtensions{.pdf,.jpg,.png}
\usepackage{mdframed}
\usepackage{xcolor}
\usepackage{hyperref}
\usepackage{epsfig}
\usepackage{float}
\usepackage{array}
\usepackage{multirow}
\usepackage{mathtools}
\usepackage{etoolbox}
\usepackage{nomencl}
\usepackage{algorithm}
\usepackage{algorithmicx}
\usepackage{algpseudocode}
\usepackage{enumerate}
\usepackage{amsmath}
\usepackage{amssymb}
\usepackage{amsthm}
\usepackage{balance}
\usepackage{soul}
\soulregister\cite7 %
\soulregister\citep7 %
\soulregister\citet7 %
\soulregister\ref7 %
\soulregister\pageref7 %
\usepackage{xcolor}
\definecolor{myred}{RGB}{205 38 38}

\renewcommand\nomgroup[1]{%
	\item[\normalsize\itshape\bfseries
	\ifstrequal{#1}{I}{Abrreviations}{%
		\ifstrequal{#1}{P}{Variables}{%
			\ifstrequal{#1}{N}{Parameters}{%
				\ifstrequal{#1}{X}{Sets}{}}}}]%
}   
\makenomenclature
\begin{document}
\title{
Analytical Large-Signal Modeling of Inverter-based Microgrids with Koopman Operator Theory for Autonomous Control}
\author{Zixiao~Ma,~\IEEEmembership{Member,~IEEE,}
	and Zhaoyu Wang,~\IEEEmembership{Senior Member,~IEEE}
\thanks{Z. Ma and Z. Wang are with the Department of Electrical and Computer Engineering, Iowa State University, Ames, IA 50011, USA (email: zma@iastate.edu; wzy@iastate.edu). (\emph{Corresponding author: Zhaoyu Wang})}
	}
\maketitle
\begin{abstract}
The microgrid (MG) plays a crucial role in the energy transition, but its nonlinearity presents a significant challenge for large-signal power systems studies in the electromagnetic transient (EMT) time scale. In this paper, we develop a large-signal linear MG model that considers the detailed dynamics of the primary and zero-control levels based on the Koopman operator (KO) theory. Firstly, a set of observable functions is carefully designed to capture the nonlinear dynamics of the MG. The corresponding linear KO is then analytically derived based on these observables, resulting in the linear representation of the original nonlinear MG with observables as the new coordinate. The influence of external input on the system dynamics is also considered during the derivation, enabling control of the MG. We solve the voltage control problem using the traditional linear quadratic integrator (LQI) method to demonstrate that textbook linear control techniques can accurately control the original nonlinear MG via the developed KO linearized MG model. Our proposed KO linearization method is generic and can be easily extended for different control objectives and MG structures using our analytical derivation procedure. We validate the effectiveness of our methodology through various case studies.
\end{abstract}

\IEEEpeerreviewmaketitle

\section{Introduction}
\IEEEPARstart{M}{icrogrids} (MGs) are localized small-scale power systems with the integration of various distributed energy resources (DERs) such as solar panels, wind turbines, or generators to provide electricity to local consumers \cite{Vasquez2010,Bidram2012,Zhang2021,Ma2021}. They are not only essential for enhancing the resilience, reliability, and efficiency of the power network, but also key to energy transition and decarbonization \cite{Chen2021}. MGs can operate autonomously or be connected to the main grid. In grid-connected mode, the MG is mainly governed by the main grid. While in islanded mode, local controls are needed to coordinate multiple DERs. 

For simplifying the controller design, MG control is usually decoupled based on different time scales \cite{Vasquez2010,Bidram2012}. Primary and zero-control levels stabilize the DERs at the fasted and lowest layer. The secondary control  eliminates the steady-state error caused by the droop characteristics. The tertiary control focuses on economic dispatching and operation scheduling in the slowest time scale. For the secondary control level, there are two major approaches. One assumes that the zero-control level can always guarantee stability and provide fast and accurate reference tracking performance so that its dynamic model can be reduced \cite{Cui2022}. This approach significantly increases the scalability of secondary control and enables large-scale system analysis. However, it inevitably results in the loss of the faster electromagnetic transient (EMT) \cite{Pogaku2007,Rasheduzzaman2014}. Moreover, large disturbances such as data loss, outliers, time delays, etc are possible to happen in the feedback channel or actuator and  result in an inappropriate secondary control signal that finally deteriorates the stability of the MG \cite{Shafiee2014}. Therefore, another approach is to design the secondary controller with consideration of detailed dynamics of primary and zero-control levels in the EMT time scale \cite{Bidram2013c,Bidram2014b}. Such an approach can capture more fast dynamics and yields a more reliable control strategy, nonetheless, the consideration of these dynamics considerably increases the system order as well as complexifies the nonlinearity of MGs \cite{Bidram2014b}.

Control of inverter-based MGs based on a nonlinear EMT model has been widely studied over the past decade \cite{Bidram2013c,Bidram2014b,Du2022}. However, controller design for nonlinear systems is usually case-by-case and can hardly be generalized to cope with different situations, such as time-delays \cite{Shafiee2014}, uncertainties \cite{Lai2019,Lai2020}, constraints \cite{Maulik2019}, etc. Thus, some studies sort to small-signal MG models based on linearization around an equilibrium point \cite{Pogaku2007,Rasheduzzaman2014}. With these models, one can use spectral tools to easily analyze the linear dynamics of MGs and adopt textbook linear control techniques to achieve various control objectives \cite{Ma2023}. However, the results obtained with small-signal models are only valid within a neighborhood around the selected equilibrium.

Recently, the Koopman operator (KO) prevails as an effective linearization method that can accurately capture large-signal nonlinear dynamics. The essential idea is that a nonlinear dynamical system can be represented by an infinite-dimensional \textit{linear} operator on a Hilbert space of vector-valued observable functions of system states \cite{koopman1931hamiltonian}. The existing KO identification approaches can be classified into numerical and analytical ones. In \textit{numerical} methods, a finite set of observable functions will be firstly designed based on the knowledge of dynamical system nonlinearity. Then, the KO will be identified using the system state's measurement data pairs of snapshots as it evolves in time. Representative methods include dynamic mode decomposition (DMD) \cite{Saldana2017,Kandaperumal2022} and its extensions, such as extended DMD (EDMD) \cite{williams2015data}, and extended DMD with control (EDMDc) \cite{korda2018linear}, etc. Especially from the MG control perspective, the KO is applied to the secondary control problem of MG in \cite{toro2023data,gong2023}. Five observable functions are initiated and the KO is estimated by the EDMDc method with the assumption that the droop gains are known by the secondary controller. The assumption on the knowledge of local controllers is further relaxed and an enhanced observer Kalman filter to optimally identify the Koopman operator is proposed in \cite{gong2023novel}. The proposed approaches well fit the studied two-dimensional state-space model, nonetheless, they cannot capture the faster dynamics in the EMT time scale since the zero-control level is not considered. To extend such a numerical method to the MGs modeled with EMT, more observable functions need to be carefully designed. Significantly, an \textit{exponentially increased volume of data pairs} is required for the numerical methods to produce an accurate estimation of the KO. 

Another way to apply KO theory to \textit{high-order nonlinear systems} is to use \textit{analytical} methods that rely on the choice of observable functions. If the observable functions are chosen perfectly, the nonlinear system can be represented in the lifted Hilbert space without any error. However, this is usually unachievable for most practical systems. A common strategy is to start with a set of observable functions and then expand them until the error between the nonlinear model and the KO linear model is sufficiently small \cite{Chen2020}. Analytical methods provide an explicit linear model that does not need to be re-identified for different system settings as in numerical methods. However, deriving the KO analytically usually depends on the specific nonlinear dynamics of a practical system. For instance, \cite{Chen2020} studied a nonlinear attitude control problem using the KO and selected the observables as the first $n$th-order derivatives of attitude dynamics. In \cite{arnas2021approximate}, the KO was used to generate approximate analytical solutions for the motion of a satellite orbiting a non-spherical celestial body with zonal harmonics. It showed that the KO could capture any order of zonal harmonics without changing the methodology. To our best knowledge, no existing study has applied an analytical KO derivation method to MG control problems.

This paper proposes an analytical KO-based large-signal model linearization approach for inverter-dominated islanded MGs. The approach considers the detailed dynamics of primary and zero-control levels in the EMT time scale. To capture the nonlinear dynamics of the MG, we design a set of observables meticulously. Then, a KO is derived analytically to represent the original nonlinear MG linearly with these observables as the new coordinate. To demonstrate that standard linear techniques are conveniently applicable, we solve the voltage control problem using the conventional linear quadratic integrator (LQI) method as an example. The main contributions of this paper are summarized as follows:
\begin{itemize}
    \item A novel linear EMT MG model considering dynamics of primary and zero-control levels is proposed based on the KO theory that represents the nonlinear MG linearly with a finite set of tailored observable functions. 
    \item Analytically derived KO is utilized to capture the nonlinear dynamics of the MG, thereby avoiding the need for huge data sets required by numerical approaches for high-dimensional complex nonlinear systems. Furthermore, the proposed KO-based model can be smoothly embedded into sophisticated linear control schemes.
    \item The proposed analytical KO-based model linearization methodology is generic and can be extended to other MGs with different control structures and topologies.
\end{itemize}

\section{Preliminaries}\label{Section:2}
This section introduces a widely-used nonlinear MG model that forms the foundation for deriving the KO linearized model in Section III. Additionally, the KO theory is briefly presented, with a focus on external control inputs that facilitate the use of linear control techniques.
 \subsection{MG modeling}
\begin{figure}[t!]
		\centering
		\includegraphics[width=0.95\columnwidth]{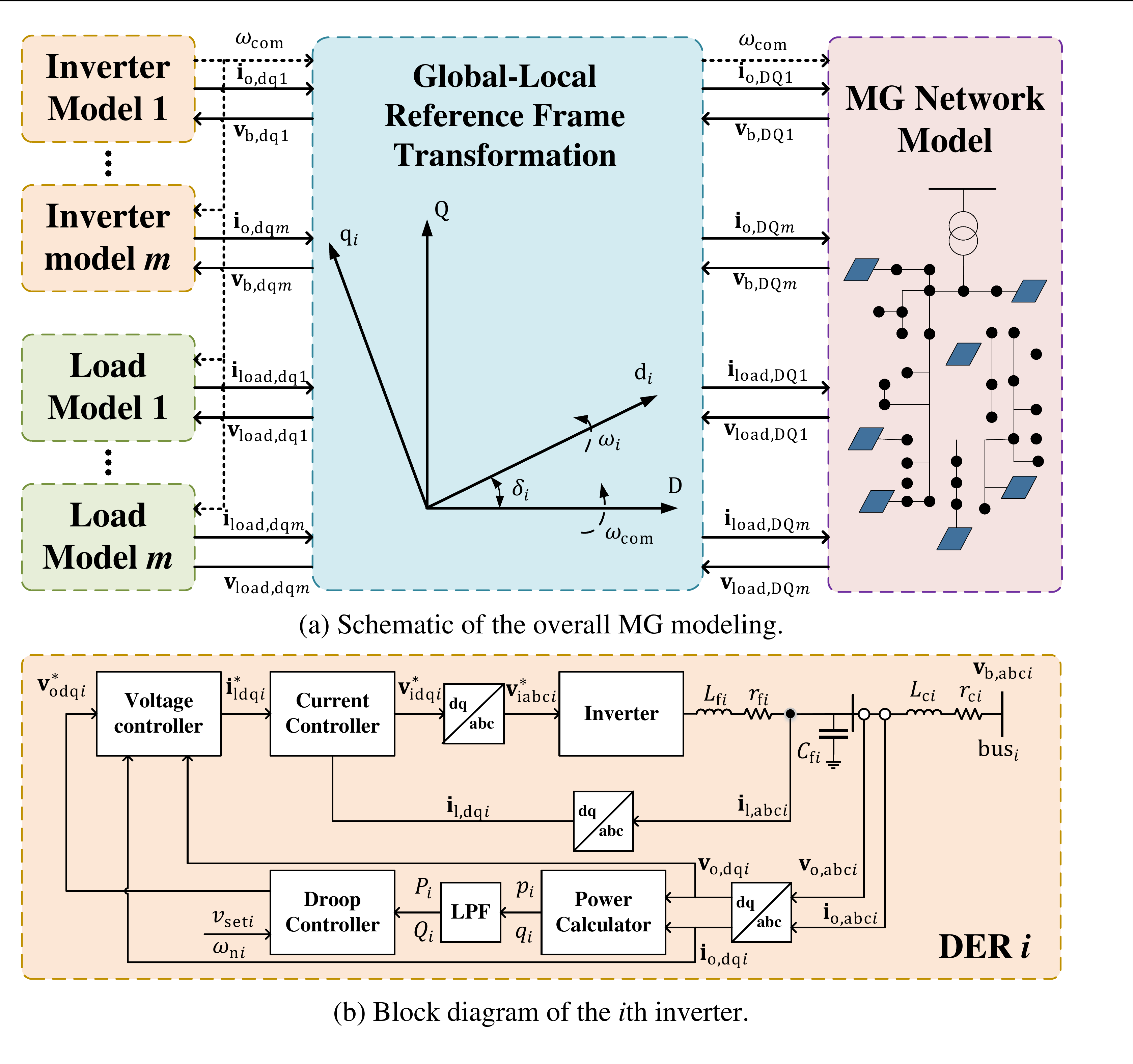}
		\caption{Overall diagram of a nonlinear MG system model.}
		\label{diagram_microgrid}
		\vspace{-1 em}
	\end{figure}

This section introduces the detailed nonlinear mathematical model of an MG based on \cite{Pogaku2007}. Figure \ref{diagram_microgrid} shows the schematic of the overall MG model that is operating in the islanded mode. The mathematical models are derived for each component of the MG in the following subsections.
\subsubsection{Power Calculation and Droop Control}
The active and reactive power produced by the system can be determined by analyzing the transformed output voltage, $v_{\rm odq}$, and current, $i_{\rm odq}$. To obtain the filtered instantaneous powers, a low-pass filter with a corner frequency of $\omega_{\rm c}$ can be utilized, which yields the following results:
\begin{subequations}\label{pq}
\begin{align}
\dot{P_i}&=-P_i\omega_{{\rm c}i}+\omega_{{\rm c}i}\left( v_{{\rm od}i}i_{{\rm od}i}+v_{{\rm oq}i}i_{{\rm oq}i}\right), \label{P} \\
\dot{Q_i}&=-Q_i\omega_{{\rm c}i}+\omega_{{\rm c}i}\left( v_{{\rm oq}i}i_{{\rm od}i}-v_{{\rm od}i}i_{{\rm oq}i}\right). \label{Q}
\end{align}	
\end{subequations}

When operating in islanded mode, a DER lacks reference inputs from the main grid, necessitating the use of droop controllers to generate its own voltage and frequency references. The process can be achieved through the following steps:
\begin{subequations}\label{droop}
\begin{align}
\omega_{i}&=\omega_{{\rm n}}-D_{{\rm P}i}P_i,\label{droop1} \\ 
v_{{\rm od}i}^\ast&=v_{{\rm set}i}-D_{{\rm Q}i}Q_i, \label{droop2}\\
v_{{\rm oq}i}^\ast&=0.
\end{align}
\end{subequations}
where $\omega_{{\rm n}}$ and $v_{{\rm set}i}$ are nominal frequency and voltage setpoints, respectively. The detailed determination of droop gains $D_{{\rm P}i}$ and $D_{{\rm Q}i}$ can be found in \cite{Bidram2014b,Pogaku2007}.

\subsubsection{Voltage and Current Controllers}
The DER output voltages and inductor currents are usually controlled via the standard proportional–integral (PI) method at the zero level. As shown below, the voltage controllers are designed to regulate the DER output voltages to their references which are generated by the droop control at the primary level:
\begin{subequations}\label{vc}
\begin{align}
\dot{\phi}_{{\rm d}i}&=v_{{\rm od}i}^\ast-v_{{\rm od}i},\label{vcq1} \\ 
i_{{\rm ld}i}^\ast&=K_{{\rm iv}i}{\phi}_{{\rm d}i}+K_{{\rm pv}i}\dot{\phi}_{{\rm d}i}+F_ii_{\rm od}-\omega_nC_{{\rm f}i}v_{\rm oq}, \label{vcd2} \\
\dot{\phi}_{{\rm q}i}&=v_{{\rm oq}i}^\ast-v_{{\rm oq}i},\label{vcq1} \\ 
i_{{\rm lq}i}^\ast&=K_{{\rm iv}i}{\phi}_{{\rm q}i}+K_{{\rm pv}i}\dot{\phi}_{{\rm q}i}+F_ii_{\rm oq}+\omega_nC_{{\rm f}i}v_{\rm od}. \label{vcq2} 
\end{align}	
\end{subequations}

The commanded voltage reference, $v_{{\rm ldq}i}^\ast$, is generated by the current controllers through the computation of the error between the reference inductor currents, $i_{{\rm ldq}i}^\ast$, and corresponding feedback measurements, $i_{{\rm ldq}i}$:
\begin{subequations}\label{cc}
	\begin{align}
	\dot{\gamma}_{{\rm d}i}&=i_{{\rm ld}i}^\ast-i_{{\rm ld}i},\label{ccd1} \\ 
 v_{{\rm id}i}^\ast&=-\omega_{{\rm n}}L_{{\rm f}i}i_{{\rm lq}i}+K_{{\rm ic}i}{\gamma}_{{\rm d}i}+K_{{\rm pc}i}\dot{\gamma}_{{\rm d}i}, \label{ccd2} \\
	\dot{\gamma}_{{\rm q}i}&=i_{{\rm lq}i}^\ast-i_{{\rm lq}i},\label{ccq1} \\ 
 v_{{\rm iq}i}^\ast&=\omega_{{\rm n}}L_{{\rm f}i}i_{{\rm ld}i}+K_{{\rm ic}i}{\gamma}_{{\rm q}i}+K_{{\rm pc}i}\dot{\gamma}_{{\rm q}i}. \label{ccq2} 
	\end{align}	
\end{subequations}

\subsubsection{LC Filters and Coupling Inductors}
By assuming that the inverter produces the demanded voltage, i.e., $v_{{\rm id}i}=v_{{\rm id}i}^\ast$, $v_{{\rm iq}i}=v_{{\rm iq}i}^\ast$, the dynamical models of LC filters and coupling inductors are as follows 
\begin{subequations}\label{lcf}
\begin{align}
\dot{i}_{{\rm ld}i}&=\left(-r_{{\rm f}i}i_{{\rm ld}i}+v_{{\rm id}i}-v_{{\rm od}i} \right)/L_{{\rm f}i}+\omega_{i}i_{{\rm lq}i} ,\label{ild} \\ 
\dot{i}_{{\rm lq}i}&=\left(-r_{{\rm f}i}i_{{\rm lq}i}+v_{{\rm iq}i}-v_{{\rm oq}i} \right)/L_{{\rm f}i}-\omega_{i}i_{{\rm ld}i} ,\label{ilq} \\
\dot{v}_{{\rm od}i}&=\left(i_{{\rm ld}i}\!-\!i_{{\rm od}i} \right)/C_{{\rm f}i}\!+\!\omega_{i}v_{{\rm oq}i},\label{vod} \\ 
\dot{v}_{{\rm oq}i}&=\left(i_{{\rm lq}i}\!-\!i_{{\rm oq}i} \right)/C_{{\rm f}i}\!-\!\omega_{i}v_{{\rm od}i}.\label{voq}\\
\dot{i}_{{\rm od}i}&=\left(-r_{{\rm c}i}i_{{\rm od}i}+v_{{\rm od}i}-v_{{\rm bd}i} \right)/L_{{\rm c}i}+\omega_{i}i_{{\rm oq}i} ,\label{iod} \\ 
\dot{i}_{{\rm oq}i}&=\left(-r_{{\rm c}i}i_{{\rm oq}i}+v_{{\rm oq}i}-v_{{\rm bq}i} \right)/L_{{\rm c}i}-\omega_{i}i_{{\rm od}i} ,\label{ioq}
\end{align}	
\end{subequations}

\subsubsection{Transforming Local Reference Frame to Global Frame}
The above mathematical model of each DER is developed in their own local $d-q$ reference frame. Suppose that the local $d-q$ reference frame of the $i$th DER is rotating at $\omega_i$ and the global $D-Q$ reference frame is rotating at $\omega_{\rm com}$. Then, we can connect each individual DER to the network by using the following rotation transformation: 
\begin{align}\label{transformation}
    \begin{bmatrix}
        x_{{\rm D}i}\\x_{{\rm Q}i}
    \end{bmatrix}=
    \begin{bmatrix}
        \cos{\delta_i}&-\sin{\delta_i}\\
        \sin{\delta_i}&\cos{\delta_i}
    \end{bmatrix}
    \begin{bmatrix}
        x_{{\rm d}i}\\x_{{\rm q}i}
    \end{bmatrix}
\end{align}
where $x$ generally represents each state variable in (\ref{pq})-(\ref{lcf}). $\delta_i$ is the difference between the global reference phase and the local one of the $i$th DER, which is defined as
\begin{align}\label{delta}
    \dot{\delta}_i=\omega_i-\omega_{\rm com}
\end{align}
For islanded MGs, the first DER is selected as the common global reference in the following derivation, i.e., $\omega_{\rm com}=\omega_1$.
\subsubsection{Network Model}
The network model is developed in the global reference frame. The dynamic model of the $i$th ($i=1,\dots,q$) line current between bus $j$ and bus $k$ is represented as follows,
\begin{subequations}\label{line}
\begin{align}
    \dot{i}_{{\rm line}i}&=(v_{{\rm bD}j}-v_{{\rm bD}k}-r_{{\rm line}i}i_{{\rm line}i})/L_{{\rm line}i}+\omega_{i}{i_{{\rm lineQ}i}},\\
    \dot{i}_{{\rm line}i}&=(v_{{\rm bQ}j}-v_{{\rm bQ}k}-r_{{\rm line}i}i_{{\rm line}i})/L_{{\rm line}i}-\omega_{i}{i_{{\rm lineD}i}}.
\end{align}
\end{subequations}
\subsubsection{Load Model}
As in \cite{Pogaku2007}, purely resistive loads and resisters and inductors (RL loads) are considered. The purely resistive loads directly follow Ohm's law without dynamics. While the $i$th ($i=1,\dots,p$) RL load can be modeled as,
\begin{subequations}\label{load}
\begin{align}
       \dot{i}_{{\rm loadD}i}&=(v_{{\rm bD}i}-R_{{\rm load}i}i_{{\rm loadD}i})/L_{{\rm load}i}+\omega_{i}{i_{{\rm loadQ}i}},\\
    \dot{i}_{{\rm loadQ}i}&=(v_{{\rm bQ}i}-R_{{\rm load}i}i_{{\rm loadQ}i})/L_{{\rm load}i}-\omega_{i}{i_{{\rm loadD}i}}.
\end{align}
\end{subequations}
The frequency is constant throughout the network, thus the dynamic equations of lines and loads can adopt $\omega_1$ derived from the first
inverter \cite{Rasheduzzaman2014}.
\subsubsection{Virtual Resistor Method}
As shown in (\ref{lcf}), (\ref{line}) and (\ref{load}), the bus voltages are treated as inputs to each subsystem, such that the influences of load perturbation could not be precisely predicted \cite{Rasheduzzaman2014}. To define the bus voltage, a virtual resistor is assumed between each bus and the ground. By selecting a sufficiently large resistance $r_{\rm n}$ for the virtual resistor, its impact on the system dynamics can be negligible. Then, the bus voltage connecting the inverters, loads and the network can be defined as 
\begin{subequations}\label{vrm}
\begin{align}
    v_{{\rm bD}i}=r_{\rm n}(i_{{\rm oD}i}-i_{{\rm loadD}i}+\sum_{j=1}^{N}i_{{\rm lineD}i,j}),\\
    v_{{\rm bQ}i}=r_{\rm n}(i_{{\rm oQ}i}-i_{{\rm loadQ}i}+\sum_{j=1}^{N}i_{{\rm lineQ}i,j})
\end{align}
\end{subequations}
where $N$ is the number of lines connected to bus $i$. Care should be taken on the direction of line currents in the last term of (\ref{vrm}). We assume the current entering the bus to be positive and the current leaving the bus to be negative.
\subsection{Compact Nonlinear Model of an MG for Voltage Control}\label{Section:2.1}
For the ease of deriving KO for the MG system, we stack up the state variables to form a compact state space model. From the viewpoint of voltage control, an inverter-based islanded MG with $m$ DERs, $p$ RL loads, and $q$ lines can be represented as follows:
	\begin{align}\label{system_o}
	{\dot{{\mathbf{x}}}(t)}&=\mathbf{f}({\mathbf{x}(t)},\mathbf{u}(t)),
	\end{align}
where ${\mathbf{x}}=\left[{\mathbf{x}_{{\rm inv}1}^{\top}},\dots,{\mathbf{x}_{{\rm inv}m}^{\top}}, {\mathbf{x}_{{\rm line}1}^{\top}},\dots,{\mathbf{x}_{{\rm line}q}^{\top}}, {\mathbf{x}_{{\rm load}1}^{\top}},\dots,\right.\\\left.{\mathbf{x}_{{\rm load}p}^{\top}}\right]^{\top}$ is the state vector of inverters, lines and loads; ${\mathbf{x}}_{{\rm inv}i}=\left[\delta_i, P_i, Q_i, \phi_{{\rm d}i}, \phi_{{\rm q}i}, \gamma_{{\rm d}i}, \gamma_{{\rm q}i}, i_{{\rm ld}i}, i_{{\rm lq}i}, v_{{\rm od}i}, v_{{\rm oq}i}, i_{{\rm od}i}, i_{{\rm oq}i}\right]^{\top}, i=1,\dots,m$, denotes the state variables of the $i^{\rm th}$ DER;  $\mathbf{x}_{{\rm line}i}=\left[i_{{\rm lineD}i}, i_{{\rm lineQ}i}\right]^{\top}, i=1,\dots,q$, are the currents of the $i^{\rm th}$ line; $\mathbf{x}_{{\rm load}i}=\left[i_{{\rm loadD}i}, i_{{\rm loadQ}i}\right]^{\top}, i=1,\dots,p$, are the currents of the $i^{\rm th}$ load; $\mathbf{u}=\left[v_{{\rm set}1},\dots,v_{{\rm set}m}\right]^{\top}$ denotes the voltage control signal to be designed. Denoting $n=13m+2p+2q$, $\mathbf{f}: \mathbb{R}^{n}\times \mathbb{R}^m\to\mathbb{R}^{n}$ is the state function describing the nonlinear system dynamics. This {high-dimensional} dynamic model represents the detailed transient dynamics of the whole MG in the EMT time scale, thus facilitating fast dynamical analysis and control.

\subsection{Brief Introduction of Koopman Operator Theory}\label{Section:2.1}
The MG system described in (\ref{system_o}) comprehensively models the primary and zero-control levels, resulting in a high-dimensional nonlinear system. Despite the increasing importance of stability analysis and controller design for dynamical systems, the system's nonlinearity presents a significant challenge for comprehensive analysis. Traditional nonlinear control methods, in particular, exhibit low generality and require complex potential function designs. From a practical standpoint, it is crucial to develop an accurate large-signal linearized MG model that bridges existing mature linear control methods and the nonlinear MG system.

\begin{figure}[t!]
		\centering
		\includegraphics[width=0.85\columnwidth]{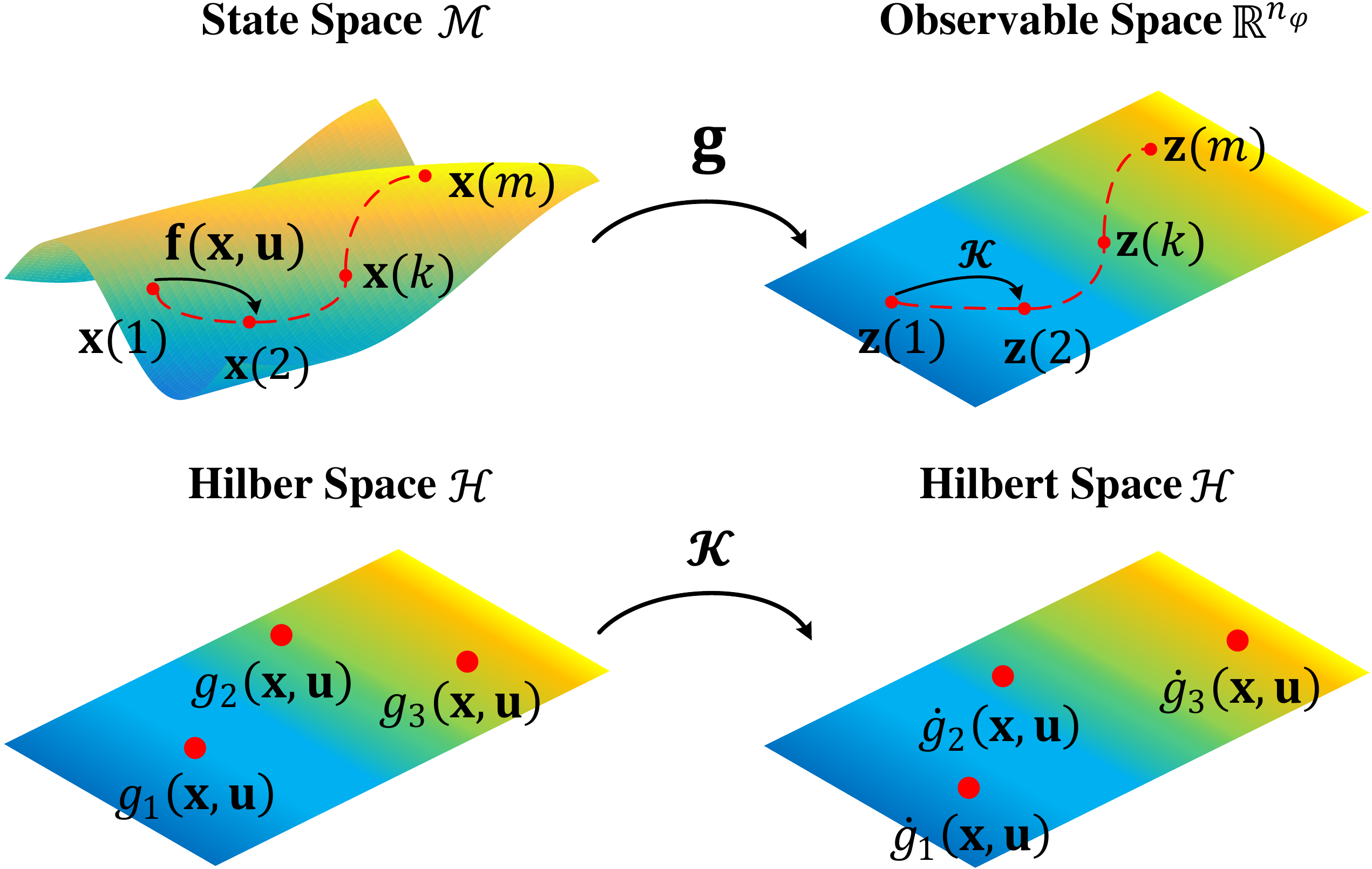}
		\caption{Illustration of the KO theory. The upper row illustrates that a dynamic system can be measured by an infinite set of observable functions $\mathbf{g}$. The lower row explains that the KO, $\mathcal{K}$, describes the dynamical evolution of the observation of the state and input, $\mathbf{g}(\mathbf{x},\mathbf{u})$, in a linear manner.}
		\label{koopman_illu}
		\vspace{-1 em}
	\end{figure}

The KO theory has gained considerable attention in nonlinear control theory and application as an effective linearization method that can accurately capture large-signal nonlinear dynamics. The fundamental concept of KO theory is to represent a nonlinear system as an infinite-dimensional linear operator on a Hilbert space of vector-valued observable functions $\mathbf{g}$ of system states. Recalling the MG system model (\ref{system_o}), where $\mathbf{x}$ and $\mathbf{u}$ evolve on smooth manifolds $\mathcal{M}$ and $\mathcal{N}$, respectively, we define the \textit{observable vector} $\mathbf{z}=\mathbf{g}(\mathbf{x},\mathbf{u}):\mathcal{M}\times\mathcal{N}\to\mathbb{R}^{N}$. 
Then, with an infinite-dimensional linear operator acting on the observable functions, the system dynamics of (\ref{system_o}) can be described linearly in this Hilbert space, i.e., 
\begin{align}\label{Koopman_def}
    \mathcal{K}\mathbf{g}(\mathbf{x},\mathbf{u})\nonumber&=\frac{d\mathbf{g}(\mathbf{x},\mathbf{u})}{dt}\\
    &=f_1\frac{\partial \mathbf{g}}{\partial x_1}+\dots+f_n\frac{\partial \mathbf{g}}{\partial x_n}+\dot{u}_1\frac{\partial \mathbf{g}}{\partial u_1}+\dots+\dot{u}_m\frac{\partial \mathbf{g}}{\partial u_m}.
\end{align}
where $\mathbf{x}=[x_1,\dots,x_n]$ and $\mathbf{u}=[u_1,\dots,u_m]$.
 In Eq. (\ref{Koopman_def}), we follow the assumption in \cite{gong2023} that the control signals influence the state evolution, but they are not evolving dynamically, i.e., $\dot{\mathbf{u}}=\mathbf{0}$. The above equation (\ref{Koopman_def}) indicates that the KO intrinsically describes the dynamical evolution of the observation of the state and input $\mathbf{g}(\mathbf{x},\mathbf{u})$ in a linear manner as illustrated in Fig. \ref{koopman_illu}. Therefore, it sheds light on analyzing the system dynamics with spectral methods and design controllers with the existing general linear control methodologies for nonlinear systems (\ref{system_o}) in the KO-oriented linear space. 

From a practical engineering perspective, it is important to note that an infinite-dimensional system is not feasible. Therefore, the key to utilizing KO theory lies in identifying an appropriate set of finite-dimensional observables and the corresponding KO that captures the primary dynamics in the Hilbert space. In the following section, we develop a KO linearized MG model with finite-dimensional observables using an analytical approach.

\section{Derivation of KO Linearized MG Model}\label{Section:3}
In this section, we present an analytical method to develop a KO linearized model of the MG system (\ref{system_o}) in the EMT time-scale, which is proposed for the first time. The derivation process involves several steps. First, assumptions are made to eliminate the nonlinearities that have negligible impact on the model accuracy. Second, we rearrange the elements in $\mathbf{x}$ to separate the linear and nonlinear terms of the system (\ref{system_o}). Third, the KO theory is applied to eliminate the nonlinear terms by designing and extending tailored observable functions. The selection of appropriate observable functions is crucial to ensure the stabilizability of the new linear system for MG voltage control. Finally, we present the KO linearized model in a concise form.
\subsection{Assumptions}
To simplify the derivation, we make some reasonable assumptions: 1) Since DER 1 is chosen as the common global reference, the difference angle between its global and local reference frame is $\delta_1=0$ with a zero initial value based on Eq. (\ref{delta}). Therefore, around the equilibrium, $\delta_i$ are small and we can approximate that $\sin\delta_i\approx\delta_i$ and $\cos\delta_i\approx1$; 2) Since the $P-\omega$ droop gain is minuscule, we assume $\omega_i\approx\omega_n$ \textit{only} in the coupling inductor terms in LC filters (\ref{lcf}) and line currents (\ref{line}). 3) More common resistive loads are considered in the following derivation to reduce the load dynamics. We rigorously test the model error caused by these assumptions in Section \ref{modelerror} under different conditions. The result shows that these assumptions are valid and acceptable.
\subsection{Separating Linear and Nonlinear Subsystems}
Based on the above assumptions, some state variables exhibit linear dynamics with respect to the system state $\mathbf{x}$ from Eq. (\ref{pq}) to Eq. (\ref{vrm}). We simplify the derivation by directly extracting and incorporating these linear equations into the final KO linearized model and addressing the remaining nonlinear dynamics with the KO.
\subsubsection{Linear subsystems} Define state vector whose dynamics linearly depends on $\mathbf{x}$ as 
\begin{align}
    \mathbf{x}_{{\rm L}i}=\left[\delta_i, \phi_{{\rm d}i}, \phi_{{\rm q}i}, \gamma_{{\rm d}i}, \gamma_{{\rm q}i}, i_{{\rm ld}i}, i_{{\rm lq}i}, v_{{\rm od}i}, v_{{\rm oq}i}\right]^{\top},\nonumber\\ i=2,\dots,m.
\end{align}

Since DER 1 is selected as the common reference, it has $\sin\delta_1=0$, $\cos\delta_1=1$ with $\delta_1(0)=0$. Then, for DER 1, the nonlinearities caused by frame transformation (\ref{transformation}) for $v_{{\rm bd}1}$ and $v_{{\rm bq}1}$ are eliminated, such that (\ref{iod})-(\ref{ioq}) become linear equations with $i=1$, i.e.,
\begin{align}
    \mathbf{x}_{{\rm L}1}=\left[\phi_{{\rm d}1}, \phi_{{\rm q}1}, \gamma_{{\rm d}1}, \gamma_{{\rm q}1}, i_{{\rm ld}1}, i_{{\rm lq}1}, v_{{\rm od}1}, v_{{\rm oq}1}, i_{{\rm od}1}, i_{{\rm oq}1}\right]^{\top}.
\end{align}

The state-space model with respect to $\mathbf{x}_{{\rm L}}=[\mathbf{x}_{{\rm L}1}^\top,\dots,\mathbf{x}_{{\rm L}m}^\top]^\top$ is derived respectively as
\begin{align}
    \dot{\mathbf{x}}_{{\rm L}1}&=\mathcal{A}_{{\rm inv}1}\mathbf{x}_{{\rm L}1}+\mathcal{A}_{1}[Q_1,i_{{\rm lineD}1},i_{{\rm lineQ}1}]^\top+\mathcal{B}_{1}v_{{\rm set}1},\\
    \dot{\mathbf{x}}_{{\rm L}i}&=\mathcal{A}_{{\rm inv}i}\mathbf{x}_{{\rm L}i}+\mathcal{A}_{i}[P_1,P_i,Q_i,i_{{\rm od}i},i_{{\rm oq}i}]^\top+\mathcal{B}_{i}v_{{\rm set}i}
\end{align}
where $\mathcal{A}_{{\rm inv}1}$, $\mathcal{A}_{{\rm inv}i}$, $\mathcal{A}_{1}$ and $\mathcal{A}_{i}$ are given in (\ref{Ainv1})-(\ref{Ainvi}), respectively and $\mathcal{B}_1=[1,0,K_{{\rm pv}1},0,b_{1},0,0,0,0,0]^\top$, $\mathcal{B}_i=[0,1,0,K_{{\rm pv}i},0,b_{i},0,0,0]^\top$ for $i=2,\dots,m$; moreover
\begin{align}
    a_{i,1}&=\frac{K_{{\rm pc}i}K_{{\rm pv}i}D_{{\rm Q}i}}{L_{{\rm f}i}},a_{i,2}=\frac{K_{{\rm pc}i}K_{{\rm iv}i}}{L_{{\rm f}i}},a_{i,3}=\frac{K_{{\rm ic}i}}{L_{{\rm f}i}},\nonumber\\
    a_{i,4}&=\frac{r_{{\rm f}i}+K_{{\rm pc}i}}{L_{{\rm f}i}},a_{i,5}=\frac{1+K_{{\rm pc}i}K_{{\rm pv}i}}{L_{{\rm f}i}},a_{i,6}=\frac{K_{{\rm pc}i}\omega_{\rm n}C_{{\rm f}i}}{L_{{\rm f}i}},\nonumber\\
    a_{i,7}&=\frac{K_{{\rm pc}i}F_i}{L_{{\rm f}i}},a_{i,8}=\frac{r_{{\rm c}i}}{L_{{\rm c}i}}+\frac{R_{{\rm load}i}r_{\rm n}}{L_{{\rm c}i}(r_{\rm n}+R_{{\rm load}i})},\nonumber\\
    a_{i,9}&=\frac{R_{{\rm load}i}r_{\rm n}}{L_{{\rm c}i}(r_{\rm n}+R_{{\rm load}i})},b_i=\frac{K_{{\rm pc}i}K_{{\rm pv}i}}{L_{{\rm f}i}},
    \;\text{for}\;i=1,\dots,m.\nonumber
\end{align}
    \begin{align}\label{Ainv1}
    &\mathcal{A}_{{\rm inv}1}=\\&\setlength {\arraycolsep} { 0.1 pt}\begin{bmatrix}
        0  &  0  &  0  &  0  &  0  &  0  &  -1  &  0  &  0  &  0\\
        0  &  0  &  0  &  0  &  0  &  0  &  0  &  -1  &  0  &  0\\
        K_{{\rm iv}1}  &  0  &  0  &  0  &  -1  &  0  &  -K_{{\rm pv}1}  &  -\omega_{\rm n}C_{{\rm f}1} &  F_1  &  0\\
        0  &  K_{{\rm iv}1}  &  0  &  0  &  0  &  -1  &  \omega_{\rm n}C_{{\rm f}1}  &  -K_{{\rm pv}1} &  0  &  F_1\\
        a_{1,2}  &  0  &  a_{1,3}  &  0  &  -a_{1,4}  &  0  &  -a_{1,5}  &  -a_{1,6} &  a_{1,7}  &  0\\
        0  &  a_{1,2}  &  0  &  a_{1,3}  &  0  &  -a_{1,4}  &  a_{1,6}  &  -a_{1,5} &  0  &  a_{1,7}\\
        0  &  0  &  0  &  0  &  \frac{1}{C_{{\rm f}1}}  &  0  &  0  &  \omega_{\rm n}  &  -\frac{1}{C_{{\rm f}1}}   &  0\\
        0  &  0  &  0  &  0  &  0  &  \frac{1}{C_{{\rm f}1}}  &  -\omega_{\rm n}  &  0  &  0  &  -\frac{1}{C_{{\rm f}1}} \\
        0  &  0  &  0  &  0  &  0  &  0  &  \frac{1}{L_{{\rm c}1}}  &  0  &  -a_{1,8}  &  \omega_{\rm n}\\
        0  &  0  &  0  &  0  &  0  &  0  &  0  &  \frac{1}{L_{{\rm c}1}}  &  -\omega_{\rm n}  &  -a_{1,8}        
    \end{bmatrix}\nonumber
    \end{align}
    \begin{align}
    \!\!\!\!\!\mathcal{A}_{{\rm inv}i}\!=\!\setlength {\arraycolsep} { 0.5 pt}\begin{bmatrix}
        0&0  &  0  &  0  &  0  &  0  &  0  &  0  &  0    \\
        0&0  &  0  &  0  &  0  &  0  &  0  &  -1  &  0  \\
        0&0  &  0  &  0  &  0  &  0  &  0  &  0  &  -1  \\
        0&K_{{\rm iv}i}  &  0  &  0  &  0  &  -1  &  0  &  -K_{{\rm pv}i}  &  -\omega_{\rm n}C_{{\rm f}i} \\
        0&0  &  K_{{\rm iv}i}  &  0  &  0  &  0  &  -1  &  \omega_{\rm n}C_{{\rm f}i}  &  -K_{{\rm pv}i} \\
        0&a_{i,2}  &  0  &  a_{i,3}  &  0  &  -a_{i,4}  &  0  &  -a_{i,5}  &  -a_{i,6} \\
        0&0  &  a_{i,2}  &  0  &  a_{i,3}  &  0  &  -a_{i,4}  &  a_{i,6}  &  -a_{i,5} \\
        0&0  &  0  &  0  &  0  &  \frac{1}{C_{{\rm f}i}}  &  0  &  0  &  \omega_{\rm n} \\
        0&0  &  0  &  0  &  0  &  0  &  \frac{1}{C_{{\rm f}i}}  &  -\omega_{\rm n}  &  0    
    \end{bmatrix}
    \end{align}
        \begin{align}
    \mathcal{A}_{1}=\begin{bmatrix}
        -D_{{\rm Q}1}  &  0  &  0  \\
        0  &  0  &  0 \\
         -K_{{\rm pv}1}D_{{\rm Q}1} &  0  &  0 \\
        0  &  0  &  0 \\
        -a_{1,1}  &  0  &  0 \\
        0  & 0  &  0  \\
        0  &  0  &  0  \\
        0  &  0  &  0 \\
        0  &  a_{1,9}  &  0  \\
        0  &  0  &  -a_{1,9}        
    \end{bmatrix}
            \end{align}
            	\vspace{-1 em}
    \begin{align}\label{Ainvi}
    \mathcal{A}_{i}=\begin{bmatrix}
        D_{{\rm P}1}&-D_{{\rm P}i}  &  0 &  0&  0 \\
        0&0&  -D_{{\rm Q}i}  &  0 &  0\\
         0&0  &  0 &  0&  0 \\
        0&0 &  -K_{{\rm pv}i}D_{{\rm Q}i}  &  F_i&  0 \\
        0&0  &  0 &  0 &  F_i \\
        0&0  &  -a_{i,1} &  a_{i,7} &  0  \\
        0&0  &  0  &  0 &  a_{i,7} \\
        0&0  &  0 &  -\frac{1}{C_{{\rm f}i}} &  0 \\
        0&0  &  0&0&-\frac{1}{C_{{\rm f}i}}         
    \end{bmatrix}
\end{align}

\subsubsection{Nonlinear subsystems (DER output power)} 
We rewrite the dynamics of active and reactive powers (\ref{pq}) as 
\begin{align}\label{zpq1}
    \underbrace{\begin{bmatrix}        
    \dot{P}_i\\\dot{Q}_i   
    \end{bmatrix}}_{\dot{\mathbf{x}}_{{\rm pq}i}}=&-\underbrace{\begin{bmatrix}        \omega_{{\rm c}i}&0\\0&\omega_{{\rm c}i}
    \end{bmatrix}}_{\mathbf{W}_{{\rm c}i}}\underbrace{\begin{bmatrix}        
    P_i\\Q_i   
    \end{bmatrix}}_{\mathbf{x}_{{\rm pq}i}}\nonumber\\&+\underbrace{\begin{bmatrix}        \omega_{{\rm c}i}&0\\0&\omega_{{\rm c}i}
    \end{bmatrix}}_{\mathbf{W}_{{\rm c}i}}\underbrace{\begin{bmatrix}        v_{{\rm od}i}&v_{{\rm oq}i}\\v_{{\rm oq}i}&-v_{{\rm od}i}
    \end{bmatrix}}_{\mathbf{V}_{{\rm o}i}}\underbrace{\begin{bmatrix}        
    i_{{\rm od}i}\\i_{{\rm oq}i}   
    \end{bmatrix}}_{\mathbf{I}_{{\rm o}i}}\triangleq\underbrace{\begin{bmatrix}        
    z_{i,1}\\z_{i,2}  
    \end{bmatrix}}_{\mathbf{z}_{i,1}}.
\end{align}
In (\ref{zpq1}), $\mathbf{z}_{i,1}$ is a designed observable vector. For the control perspective, we take the second derivative of $\mathbf{z}_{i,1}$ until the control signal $\mathbf{u}$ appears in the second derivative of DER output voltage $\Ddot{v}_{{\rm od}i}$. The derivation process is as follows,
\begin{align}
    \dot{\mathbf{z}}_{i,1}&=-\mathbf{W}_{{\rm c}i}\mathbf{z}_{i,1}+\mathbf{W}_{{\rm c}i}(\dot{\mathbf{V}}_{{\rm o}i}\mathbf{I}_{{\rm o}i}+\dot{\mathbf{V}}_{{\rm o}i}\dot{\mathbf{I}}_{{\rm o}i})\triangleq\mathbf{z}_{i,2},\\
    \dot{\mathbf{z}}_{i,2}&=-\mathbf{W}_{{\rm c}i}\mathbf{z}_{i,2}+\mathbf{W}_{{\rm c}i}(\Ddot{\mathbf{V}}_{{\rm o}i}\mathbf{I}_{{\rm o}i}+2\dot{\mathbf{V}}_{{\rm o}i}\dot{\mathbf{I}}_{{\rm o}i}+\mathbf{V}_{{\rm o}i}\Ddot{\mathbf{I}}_{{\rm o}i}).\label{zpq2nd}
\end{align}
Define the second term at the right-hand side of (\ref{zpq2nd}) as $\mathbf{U}_{{\rm pq}i}$:
\begin{align}\label{upqi}
    \mathbf{U}_{{\rm pq}i}&=\mathbf{W}_{{\rm c}i}(\Ddot{\mathbf{V}}_{{\rm o}i}\mathbf{I}_{{\rm o}i}+2\dot{\mathbf{V}}_{{\rm o}i}\dot{\mathbf{I}}_{{\rm o}i}+\mathbf{V}_{{\rm o}i}\Ddot{\mathbf{I}}_{{\rm o}i})\nonumber\\
    &=\mathbf{W}_{{\rm c}i}\left(\begin{bmatrix}
        \Ddot{v}_{{\rm oq}i}i_{{\rm oq}i}\\\Ddot{v}_{{\rm oq}i}i_{{\rm od}i}
    \end{bmatrix}+2\dot{\mathbf{V}}_{{\rm o}i}\dot{\mathbf{I}}_{{\rm o}i}+\mathbf{V}_{{\rm o}i}\Ddot{\mathbf{I}}_{{\rm o}i}+\begin{bmatrix}
        \Ddot{v}_{{\rm od}i}i_{{\rm od}i}\\-\Ddot{v}_{{\rm od}i}i_{{\rm oq}i}
    \end{bmatrix}\right)\nonumber\\
    &\triangleq\mathbf{f}_{{\rm pq}i}(\mathbf{x})+\mathcal{B}_{{\rm pq}i}\mathbf{u}
\end{align}
where $\mathbf{f}_{{\rm pq}i}(\mathbf{x})$ is a nonlinear vector-valued function of $\mathbf{x}$ that can be extracted by substracting $\mathcal{B}_{{\rm pq}i}\mathbf{u}$ from $\mathbf{U}_{{\rm pq}i}$ and 
\begin{align*}
    \mathcal{B}_{{\rm pq}i}=\begin{bmatrix}
        \frac{b_i\omega_{{\rm c}i}i_{{\rm od}i}}{C_{{\rm f}i}}&0&0\\
        -\frac{b_i\omega_{{\rm c}i}i_{{\rm oq}i}}{C_{{\rm f}i}}&0&0
    \end{bmatrix}
\end{align*}

In conclusion, we define the observable vector for the nonlinear subsystems with respect to DER output power as 
\begin{align}
    \mathbf{z}_{{\rm pq}i}=\left[\mathbf{x}_{{\rm pq}i}^\top,\mathbf{z}_{i,1}^\top,\mathbf{z}_{i,2}^\top\right]^\top,\;i=1,\dots,m.
\end{align}
\subsubsection{Nonlinear subsystems (currents of DERs and network)} 
\begin{figure}[t!]
		\centering
		\includegraphics[width=0.95\columnwidth]{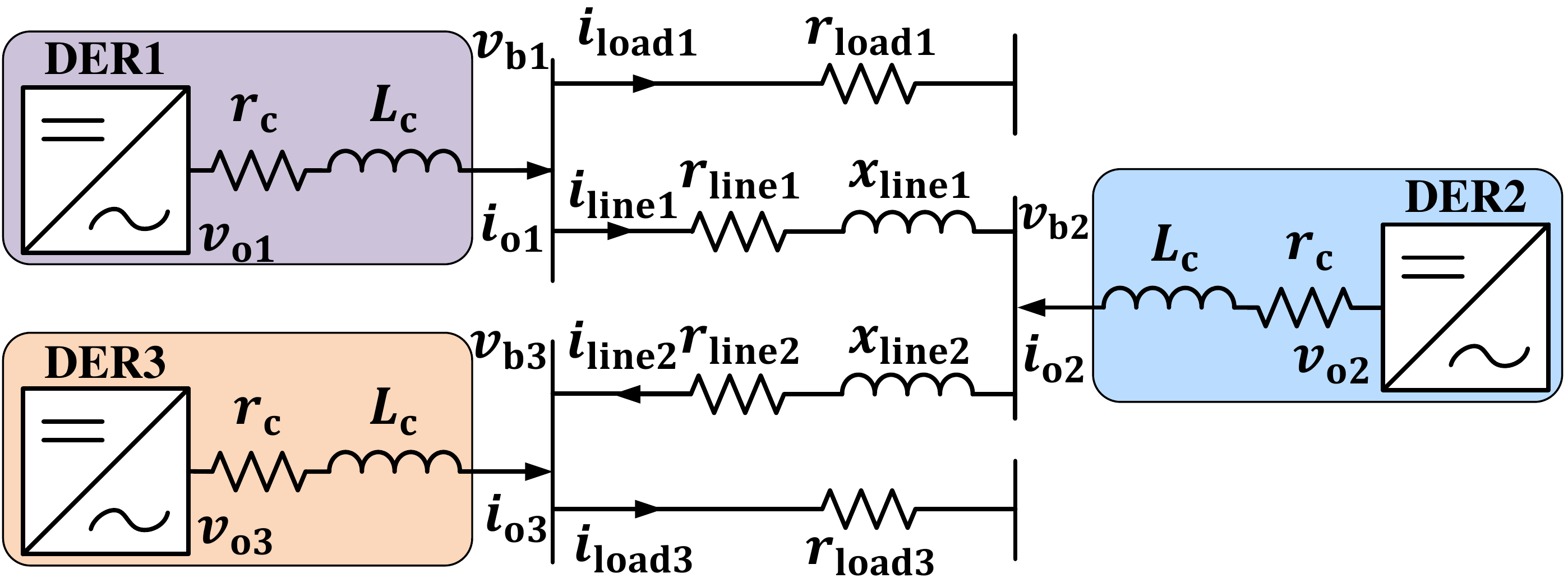}
		\caption{Diagram of the test MG system.}
		\label{testfig}
						\vspace{-1 em}
	\end{figure}

Since the DER output currents are coupled with the network currents, we handle them together and define
 \begin{align}
     \mathbf{x}_{\rm net}&=\left[i_{{\rm od}i}, i_{{\rm oq}i},i_{{\rm lineD}j}, i_{{\rm lineQ}j}\right]^{\top},\nonumber\\i&=2,\dots,m,\;j=1,\dots,q.
 \end{align}
Then, from (\ref{iod})-(\ref{vrm}), we rewrite the state equations as
\begin{align}
    \dot{\mathbf{x}}_{\rm net}=\mathcal{A}_{\rm net}\mathbf{x}_{\rm net}+\mathbf{H}\mathbf{\upxi}+\mathbf{D}\mathbf{x}_{\rm net}\triangleq\mathbf{z}_{{\rm net}1},
\end{align}
The positions of elements in $\mathcal{A}_{\rm net}$, $\mathbf{H}$, and $\mathbf{D}$ depend on the topology of the MG. To illustrate the derivation, we take a test system shown in Fig. \ref{testfig} as an example. Then, $\mathbf{x}_{\rm net}=\left[i_{{\rm od}2}, i_{{\rm oq}2},i_{{\rm od}3}, i_{{\rm oq}3},i_{{\rm lineD}1}, i_{{\rm lineQ}1},i_{{\rm lineD}2}, i_{{\rm lineQ}2}\right]^{\top}$, $\mathbf{\upxi}=\left[i_{{\rm od}1}, i_{{\rm oq}1},v_{{\rm od}2}, v_{{\rm oq}2},v_{{\rm od}3}, v_{{\rm oq}3}\right]^{\top}$ and the matrices are given as follows,
\begin{align}
    \mathcal{A}_{\rm net}\!=\!\setlength {\arraycolsep} { 0.1 pt}\begin{bmatrix}
        -a_{2,8}&\omega_{\rm n}&0&0&-a_{2,9}&0&a_{2,9}&0\\
        -\omega_{\rm n}&-a_{2,8}&0&0&0&a_{2,9}&0&-a_{2,9}\\
        0&0&-a_{3,8}&\omega_{\rm n}&0&0&-a_{3,9}&0\\
        0&0&-\omega_{\rm n}&-a_{3,8}&0&0&0&a_{3,9}\\
        -\frac{r_{\rm n}}{L_{{\rm line}1}}&0&0&0&-a_{10}&\omega_{\rm n}&\frac{r_{\rm n}}{L_{{\rm line}1}}&0\\
        0&-\frac{r_{\rm n}}{L_{{\rm line}1}}&0&0&-\omega_{\rm n}&-a_{10}&0&\frac{r_{\rm n}}{L_{{\rm line}1}}\\
        \frac{r_{\rm n}}{L_{{\rm line}2}}&0&-a_{13}&0&\frac{r_{\rm n}}{L_{{\rm line}2}}&0&-a_{11}&\omega_{\rm n}\\
        0&\frac{r_{\rm n}}{L_{{\rm line}2}}&0&-a_{13}&0&\frac{r_{\rm n}}{L_{{\rm line}2}}&-\omega_{\rm n}&-a_{11}
    \end{bmatrix}
\end{align}
    		\vspace{-1 em}
\begin{align}
    \mathbf{H}=\begin{bmatrix}
        0&0&\frac{1}{L_{{\rm c}2}}&0&0&0\\
       0&0&0&\frac{1}{L_{{\rm c}2}}&0&0\\
        0&0&0&0&\frac{1}{L_{{\rm c}3}}&0\\
        0&0&0&0&0&\frac{1}{L_{{\rm c}3}}\\
        a_{12}&0&0&0&0&0\\
        0&a_{12}&0&0&0&0\\
        0&0&0&0&0&0\\
        0&0&0&0&0&0
    \end{bmatrix}
\end{align}
	\vspace{-1 em}
\begin{align}
    \!\!\mathbf{D}\!\!=\!\!\setlength {\arraycolsep} { 0.02 pt}\begin{bmatrix}
        0&0&0&0&0&-a_{2,9}\delta_2&0&a_{2,9}\delta_2\\
        0&0&0&0&a_{2,9}\delta_2&0&-a_{2,9}\delta_2&0\\
        0&0&0&0&0&0&0&-a_{3,9}\delta_3\\
        0&0&0&0&0&0&a_{3,9}\delta_3&0\\
        0&\frac{r_{\rm n}\delta_2}{L_{{\rm line}1}}&0&0&0&0&0&0\\
        \frac{-r_{\rm n}\delta_2}{L_{{\rm line}1}}&0&0&0&0&0&0&0\\
        0&\frac{-r_{\rm n}\delta_2}{L_{{\rm line}2}}&0&a_{13}\delta_3&0&0&0&0\\
        \frac{r_{\rm n}\delta_2}{L_{{\rm line}2}}&0&-a_{13}\delta_3&0&0&0&0&0\\
    \end{bmatrix}
\end{align}
where the parameters $a_{10}$ to $a_{13}$ are defined as
\begin{align*}
    a_{10}&=\frac{r_{{\rm line}1}+r_{\rm n}}{L_{{\rm line}1}}+\frac{R_{{\rm load}1}r_{\rm n}}{L_{{\rm line}1}(R_{{\rm load}1}+r_{\rm n})},\\
      a_{11}&=\frac{r_{{\rm line}2}+r_{\rm n}}{L_{{\rm line}2}}+\frac{R_{{\rm load}3}r_{\rm n}}{L_{{\rm line}2}(R_{{\rm load}3}+r_{\rm n})},\\
    a_{12}&=\frac{R_{{\rm load}1}r_{\rm n}}{L_{{\rm line}1}(R_{{\rm load}1}+r_{\rm n})},a_{13}=\frac{R_{{\rm load}3}r_{\rm n}}{L_{{\rm line}2}(R_{{\rm load}3}+r_{\rm n})}
\end{align*}

For the control purpose, we take the second derivative of $\mathbf{z}_{{\rm net}1}$ until the control signal $\mathbf{u}$ appears in the second derivative of $\Ddot{v}_{{\rm od}i}$ in $\Ddot{\mathbf{\upxi}}$. The derivation process is as follows,
\begin{align}
    \!\!\!\!\dot{\mathbf{z}}_{{\rm net}1}&=\mathcal{A}_{\rm net}\mathbf{z}_{{\rm net}1}+\mathbf{H}\dot{\mathbf{\upxi}}+\dot{\mathbf{D}}\mathbf{x}_{\rm net}+\mathbf{D}\mathbf{z}_{{\rm net}1}\triangleq\mathbf{z}_{{\rm net}2},\\
    \!\!\!\!\dot{\mathbf{z}}_{{\rm net}2}&=\mathcal{A}_{\rm net}\mathbf{z}_{{\rm net}2}+\Ddot{\mathbf{D}}\mathbf{x}_{\rm net}+2\dot{\mathbf{D}}\mathbf{z}_{{\rm net}1}+\mathbf{D}\mathbf{z}_{{\rm net}2}+\mathbf{H}\Ddot{\mathbf{\upxi}}.
\end{align}
Define the control vector $\mathbf{U}_{\rm net}$ as (\ref{unet}). Note that $\mathbf{z}_{{\rm net}1}$ and $\mathbf{z}_{{\rm net}2}$ can be represented with $\mathbf{x}$, and $\mathbf{u}$ can be extracted from $\Ddot{\mathbf{\upxi}}$, thus the control vector $\mathbf{U}_{\rm net}$ can be separated as follows,
\begin{align}\label{unet}
   \mathbf{U}_{\rm net}&=\Ddot{\mathbf{D}}\mathbf{x}_{\rm net}+2\dot{\mathbf{D}}\mathbf{z}_{{\rm net}1}+\mathbf{D}\mathbf{z}_{{\rm net}2}+\mathbf{H}\Ddot{\mathbf{\upxi}}\nonumber\\
   &=\underbrace{\Ddot{\mathbf{D}}\mathbf{x}_{\rm net}+2\dot{\mathbf{D}}\mathbf{z}_{{\rm net}1}+\mathbf{D}\mathbf{z}_{{\rm net}2}+\mathbf{H}\Ddot{\mathbf{\upxi}}^{\ast}}_{\mathbf{f}_{\rm net}(\mathbf{x})}+\mathcal{B}_{\rm net}\mathbf{u}
\end{align}
where $\Ddot{\mathbf{\upxi}}^{\ast}=\Ddot{\mathbf{\upxi}}-{\mathcal{B}}_{\rm net}\mathbf{u}$, $\mathcal{B}_{\rm net}=\mathbf{H}\bar{\mathcal{B}}_{\rm net}$ and 
\begin{align}
    \bar{\mathcal{B}}_{\rm net}=\begin{bmatrix}
        0&0&0\\
        0&0&0\\
        0&b_2&0\\
        0&0&0\\
        0&0&b_3\\
        0&0&0
    \end{bmatrix}
\end{align}
In conclusion, we define the observable vector for the nonlinear subsystems with respect to DER output currents and network as 
\begin{align}
    \mathbf{z}_{{\rm net}}=\left[\mathbf{x}_{{\rm net}}^\top,\mathbf{z}_{{\rm net}1}^\top,\mathbf{z}_{{\rm net}2}^\top\right]^\top.
\end{align}
\subsection{Overall KO Linearized MG Model}
Defining the observable vector of the overall MG system as $\mathbf{z}=\left[\mathbf{x}_{{\rm L}}^\top,\mathbf{z}_{{\rm pq}1}^\top,\dots,\mathbf{z}_{{\rm pq}m}^\top,\mathbf{z}_{\rm net}^\top\right]^\top\in\mathbb{R}^{N}$, the KO linearized model can be concluded as
\begin{subequations}\label{KO model}
    \begin{align}
    \dot{\mathbf{z}}&=\mathbf{A}\mathbf{z}+\mathbf{B}\mathbf{U},\\
    \mathbf{y}&=\mathbf{C}\mathbf{z}
\end{align}
\end{subequations}
where $\mathbf{y}=[{v}_{{\rm od}1},\dots,{v}_{{\rm od}m}]^{\top}\in\mathbb{R}^{M}$ is the output vector, which can be extracted from the state vector with matrix $\mathbf{C}$, $\mathbf{U}=\left[\mathbf{u}^\top,\mathbf{U}_{{\rm pq}1}^\top,\dots,\mathbf{U}_{{\rm pq}m}^\top,\mathbf{U}_{{\rm net}}^\top\right]$ is the lifted control input vector to be designed according to the control performance requirement. Take the system in Fig. \ref{testfig} as an example, $m=3$ and $N=70$. Then the corresponding matrices $\mathbf{A}$ and $\mathbf{B}$ are derived as below
\begin{align*}
    \mathbf{A}=\setlength {\arraycolsep} { 1.5 pt}\begin{bmatrix}
        \mathcal{A}_{{\rm inv}1}&\mathbf{0}&\mathbf{0}&\mathcal{A}_{1}^1&\mathbf{0}&\mathbf{0}&\mathcal{A}_{1}^{2,3}&\mathbf{0}&\mathbf{0}\\
        \mathbf{0}&\mathcal{A}_{{\rm inv}2}&\mathbf{0}&\mathcal{A}_{2}^1&\mathcal{A}_{2}^{2,3}&\mathbf{0}&\mathcal{A}_{2}^{4,5}&\mathbf{0}&\mathbf{0}\\
        \mathbf{0}&\mathbf{0}&\mathcal{A}_{{\rm inv}3}&\mathcal{A}_{3}^1&\mathbf{0}&\mathcal{A}_{3}^{2,3}&\mathcal{A}_{2}^{4,5}&\mathbf{0}&\mathbf{0}\\
        \mathbf{0}&\mathbf{0}&\mathbf{0}&\mathcal{A}_{\omega_{1}}&\mathbf{0}&\mathbf{0}&\mathbf{0}&\mathbf{0}&\mathbf{0}\\
        \mathbf{0}&\mathbf{0}&\mathbf{0}&\mathbf{0}&\mathcal{A}_{\omega_{2}}&\mathbf{0}&\mathbf{0}&\mathbf{0}&\mathbf{0}\\
        \mathbf{0}&\mathbf{0}&\mathbf{0}&\mathbf{0}&\mathbf{0}&\mathcal{A}_{\omega_{3}}&\mathbf{0}&\mathbf{0}&\mathbf{0}\\
        \mathbf{0}&\mathbf{0}&\mathbf{0}&\mathbf{0}&\mathbf{0}&\mathbf{0}&\mathbf{0}&\mathbf{I}_{8} &\mathbf{0}\\
        \mathbf{0}&\mathbf{0}&\mathbf{0}&\mathbf{0}&\mathbf{0}&\mathbf{0}&\mathbf{0}&\mathbf{0}&\mathbf{I}_{8} \\
        \mathbf{0}&\mathbf{0}&\mathbf{0}&\mathbf{0}&\mathbf{0}&\mathbf{0}&\mathbf{0}&\mathbf{0}&\mathcal{A}_{\rm net}
    \end{bmatrix}_{70\times 70}
\end{align*}
		\begin{figure*}[t!]
		\centering
		\includegraphics[width=1.7\columnwidth]{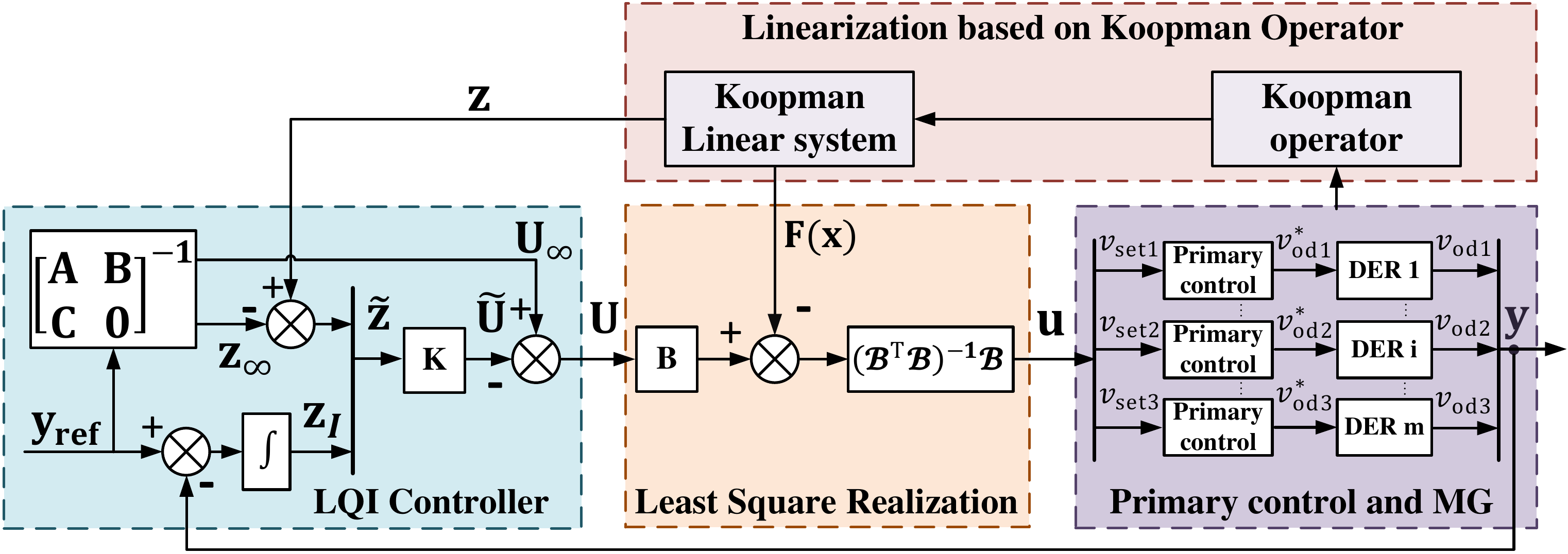}
		\caption{Closed-loop MG control system based on the KO linearized model and LQI. The LQI gain is $\mathbf{K}=\mathbf{R}^{-1}\widetilde{\mathbf{B}}^{\top}\mathbf{P}$.}
		\label{overall_control}
						\vspace{-1 em}
	\end{figure*}
\begin{align*}
    \mathbf{B}=\begin{bmatrix}
        \mathbf{B}_1&\mathbf{B}_2&\mathbf{B}_3
    \end{bmatrix}_{70\times 17}
\end{align*}

For simplification, we define the elements in $\mathbf{A}$ and $\mathbf{B}$ with MATLAB language (e.g., $\mathcal{A}_1(:,2:3)$ means the second to the third columns of matrix $\mathcal{A}_1$ and ``;'' denotes line break)  
$\mathcal{A}_{1}^1=\left[\mathbf{0}_{10\times1},\mathcal{A}_1(:,1),\mathbf{0}_{10\times4}\right]$, $\mathcal{A}_{1}^{2,3}=\left[\mathbf{0}_{10\times4},\mathcal{A}_1(:,2:3),\mathbf{0}_{10\times2}\right]$, $\mathcal{A}_{2}^1=\left[\mathcal{A}_2(:,1),\mathbf{0}_{9\times5}\right]$, 
$\mathcal{A}_{2}^{2,3}=\left[\mathcal{A}_2(:,2:3),\mathbf{0}_{9\times4}\right]$,
$\mathcal{A}_{2}^{4,5}=\left[\mathcal{A}_2(:,4:5),\mathbf{0}_{9\times6}\right]$, 
$\mathcal{A}_{3}^1=\left[\mathcal{A}_3(:,1),\mathbf{0}_{9\times5}\right]$, $\mathcal{A}_{3}^{2,3}=\left[\mathcal{A}_3(:,2:3),\mathbf{0}_{9\times4}\right]$, $\mathcal{A}_{3}^{4,5}=\left[\mathbf{0}_{9\times2},\mathcal{A}_3(:,4:5),\mathbf{0}_{9\times4}\right]$, $\mathbf{k}_{i}=\left[1,0,K_{{\rm pv}i}0,b_{i}\right]^\top$ for $i=1,2,3$. $\mathbf{B}_{1}=\left[\mathbf{ k}_{1},\mathbf{0}_{1\times65};\mathbf{0}_{1\times11},\mathbf{ k}_{2},\mathbf{0}_{1\times54};\mathbf{0}_{1\times20},\mathbf{ k}_{3},\mathbf{0}_{1\times45}\right]^\top$, $\mathbf{B}_{2}=\left[\mathbf{0}_{2\times32},\mathbf{I}_2,\mathbf{0}_{2\times36};\mathbf{0}_{2\times38},\mathbf{I}_2,\mathbf{0}_{2\times30};\mathbf{0}_{2\times44},\mathbf{I}_2,\mathbf{0}_{2\times24}\right]^\top$, $\mathbf{B}_{3}=\left[\mathbf{0}_{62\times8};\mathbf{I}_{8}\right]$, and
\begin{align*}
    \mathcal{A}_{\omega_{i}}=\begin{bmatrix}
\phantom{-}0\phantom{-}&\phantom{-}0\phantom{-}&\phantom{-}1\phantom{-}&\phantom{-}0\phantom{-}&0&0\\
        0&0&0&1&0&0\\
        0&0&0&0&1&0\\
        0&0&0&0&0&1\\
        0&0&0&0&-\omega_{i}&0\\
        0&0&0&0&0&-\omega_{i}
    \end{bmatrix}.
\end{align*}

\emph{Remark 1:} The purpose of the KO linearized model (\ref{KO model}) is to \textit{enable general linear control techniques} that are still effective for the original nonlinear system. In practical application, the lifted-dimensional controller $\mathbf{U}$ will be designed based on the auxiliary linear model (\ref{KO model}) using any general linear control methods. Then, an analytical actual control signal $\mathbf{u}$ will be obtained from $\mathbf{U}$. Finally, $\mathbf{u}$ will be applied to the original nonlinear MG system (\ref{system_o}). It should also be noted that since part of system dynamics $\mathbf{F}(\mathbf{x})$ is included in the control term $\mathbf{B}\mathbf{U}$, one should not expect stability of the original nonlinear model (\ref{system_o}) can be analyzed through the eigenvalues of $\mathbf{A}$ (assuming zero input) as usually done in small-signal models. This problem is further discussed in the case study section.
\section{Voltage Control of MG based on the KO Linearized Model}\label{Section:4}
A critical contribution of this work is that users can select any linear control methods according to their requirements on their control objectives. In this section, we use MG's voltage restoration problem as an example to demonstrate how to use the above-developed linear MG model based on the KO theory. The control objective is to eliminate the steady-state errors between the output voltages of DERs and their reference values caused by the droop characteristics \cite{Bidram2012}.

\subsection{Controller Design based on KO Linearized Model with LQI}
To achieve zero-offset voltage regulation and facilitate easy deployment, the optimal control method LQI is adopted in this section \cite{Ma2023}. 

Firstly, as shown in the very left block in Fig. \ref{overall_control}, an integrator that dynamically feeds back the integral of the offset between DER output voltages and their references is designed as follows,
\begin{align}\label{integrator}
\dot{\mathbf{z}}_{\rm I}={\mathbf{y}}_{\rm ref}-{\mathbf{y}},
\end{align}
where {${\mathbf{z}}_{\rm I}$} denotes the error dynamics of the integrator and ${\mathbf{y}}_{\rm ref}$ contains the voltage setpoints to be tracked. 
 
Then, by defining new state vector $\tilde{\mathbf{z}}\triangleq\left[{\mathbf{z}}^{\top}-{\mathbf{z}_{\infty}^{\top}}, {\mathbf{z}}_{\rm I}^{\top}\right]^{\top}$, control input vector $\widetilde{\mathbf{U}}=\left[{\mathbf{U}}-{\mathbf{U}}_{\infty}\right]$ and output offset vector $\tilde{\mathbf{y}}(k)={\mathbf{y}}(k)-{\mathbf{y}}_{\rm ref}$, the bias system is derived as follows,
\begin{subequations}\label{system_a}
\begin{align}
\dot{\tilde{\mathbf{z}}}&=\widetilde{\mathbf{A}}\tilde{\mathbf{z}}+\widetilde{\mathbf{B}}\widetilde{\mathbf{U}},\\
\tilde{\mathbf{y}}&=\widetilde{\mathbf{C}}\tilde{\mathbf{z}}
\end{align}
\end{subequations}
where the system matrices of the above-augmented system are given as 
\begin{align}
\widetilde{\mathbf{A}}=
    \begin{bmatrix}
    \mathbf{A} & \mathbf{0}\\
    -\mathbf{C} & \mathbf{0}
    \end{bmatrix}, 
    \widetilde{\mathbf{B}}=
    \begin{bmatrix}
    \mathbf{B} \\
    \mathbf{0}
    \end{bmatrix}, 
    \widetilde{\mathbf{C}}=
    \begin{bmatrix}
    \mathbf{C}  & \mathbf{0}
    \end{bmatrix}.
\end{align}

Finally, to achieve offset-free setpoint tracking, the steady-state values ${\mathbf{z}_{\infty}}$ and ${\mathbf{U}}_{\infty}$ should satisfy \begin{align}\label{steady_state}
    \begin{bmatrix}
    \mathbf{A} & \mathbf{B} \\
    \mathbf{C} & \mathbf{0}
    \end{bmatrix}\begin{bmatrix}
    {\mathbf{z}_{\infty}}\\
    {\mathbf{U}}_{\infty}
    \end{bmatrix}=
    \begin{bmatrix}
    \mathbf{0}\\
    {\mathbf{y}}_{\rm ref}
    \end{bmatrix}. 
\end{align}

Considering the following optimal performance index for the continuous-time system (\ref{system_a}),
\begin{align}
    J=\frac{1}{2}\int_{t=0}^{\infty}(\tilde{\mathbf{z}}^{\top}\mathbf{Q}\tilde{\mathbf{z}}+\widetilde{\mathbf{U}}^{\top}\mathbf{R}\widetilde{\mathbf{U}})dt,
\end{align}
where $\mathbf{Q}$ and $\mathbf{R}$ are weighting matrices. The optimal control law minimizing $J$ is derived as
\begin{align}\label{LQR_law}
\widetilde{\mathbf{U}}&=-\mathbf{R}^{-1}\widetilde{\mathbf{B}}^{\top}\mathbf{P}\tilde{\mathbf{z}},\\
\mathbf{U}&=-\mathbf{R}^{-1}\widetilde{\mathbf{B}}^{\top}\mathbf{P}\tilde{\mathbf{z}}+{\mathbf{U}}_{\infty},\label{LQR_law2}
\end{align}
where $\mathbf{P}$ is the unique positive definite solution to the following continuous-time algebraic Riccati equation 
\begin{align}
\widetilde{\mathbf{A}}^{\top}\mathbf{P}+\mathbf{P}\widetilde{\mathbf{A}}-\mathbf{P}\widetilde{\mathbf{B}}\mathbf{R}^{-1}\widetilde{\mathbf{B}}^{\top}\mathbf{P}+\mathbf{Q}=\mathbf{0}.
\end{align} 

When the bias system (\ref{system_a})-(\ref{steady_state}) is stabilized by (\ref{LQR_law}), it is equivalent that: a) the KO linearized model (\ref{KO model}) is stabilized; b) the DER output voltages of (\ref{KO model}), $\mathbf{y}$ is regulated to the setpoint $\mathbf{y}_{\rm ref}$ with zero offsets, since $\dot{\mathbf{z}}_{\rm I}={\mathbf{y}}_{\rm ref}-{\mathbf{y}}=0$.

\subsection{Recovering Lower-Dimensional Control Signal for the Original MG System from the Lifted Control Vector}
Note that the lifted control vector $\mathbf{U}\in\mathbb{R}^{M}$  of the KO linearized model (\ref{KO model}) is of higher dimensional than the control vector $\mathbf{u}\in\mathbb{R}^{m}$  of the original nonlinear MG model (\ref{system_o}). Thus, the lifted control signal $\mathbf{U}$ is not directly applicable. Since the first three elements of $\mathbf{U}$ are just $\mathbf{u}$, one can use them as the control inputs of the original MG system. However, such a choice is no longer optimal due to the loss of information of the other elements in $\mathbf{U}$. Therefore, we propose the following optimal control signal recovery method.

Denote $\mathbf{U}_{{\rm pq}}=\left[\mathbf{U}_{{\rm pq}1},\dots,\mathbf{U}_{{\rm pq}m}\right]^\top$, $\mathcal{B}_{{\rm pq}}=\left[\mathcal{B}_{{\rm pq}1},\dots,\mathcal{B}_{{\rm pq}m}\right]^\top$ and $\mathbf{f}_{{\rm pq}}=\left[\mathbf{f}_{{\rm pq}1},\dots,\mathbf{f}_{{\rm pq}m}\right]^\top$, from (\ref{upqi}) and (\ref{unet}), it has
\begin{align}
    \mathbf{B}\mathbf{U}&=\mathbf{B}_1\mathbf{u}+\mathbf{B}_2\mathbf{U}_{{\rm pq}}+\mathbf{B}_3\mathbf{U}_{{\rm net}}\nonumber\\
    &=\mathbf{B}_1\mathbf{u}+\mathbf{B}_2\left(\mathbf{f}_{{\rm pq}}(\mathbf{x})+\mathcal{B}_{{\rm pq}}\mathbf{u}\right)+\mathbf{B}_3\left(\mathbf{f}_{{\rm net}}(\mathbf{x})+\mathcal{B}_{{\rm net}}\mathbf{u}\right)\nonumber\\
    &=\underbrace{\mathbf{B}_2\mathbf{f}_{{\rm pq}}(\mathbf{x})+\mathbf{B}_3\mathbf{f}_{{\rm net}}(\mathbf{x})}_{\mathbf{F}(\mathbf{x})}+\underbrace{\left(\mathbf{B}_1+\mathbf{B}_2\mathcal{B}_{{\rm pq}}+\mathbf{B}_3\mathcal{B}_{{\rm net}}\right)}_{\mathcal{B}}\mathbf{u}
\end{align}
Notice that matrix $\mathcal{B}$ is not a square matrix such that $\mathbf{u}$ cannot be directly retrieved via $\mathcal{B}^{-1}$. Therefore, we optimally recover $\mathbf{u}$ from $\mathbf{U}$ by solving the following least square problem, 
\begin{align}
    \min \frac{1}{2}\left(\mathcal{B}\mathbf{u}-(\mathbf{B}\mathbf{U}-\mathbf{F}(\mathbf{x}))\right)^\top\left(\mathcal{B}\mathbf{u}-(\mathbf{B}\mathbf{U}-\mathbf{F}(\mathbf{x}))\right)
\end{align}
whose solution is 
\begin{align}\label{sol_least}
    \mathbf{u}=(\mathcal{B}^\top\mathcal{B})^{-1}\mathcal{B}^\top\left(\mathbf{B}\mathbf{U}-\mathbf{F}(\mathbf{x})\right)
\end{align}
substituting (\ref{LQR_law2}) into (\ref{sol_least}), the controller for original MG (\ref{system_o}) is obtained as follows
\begin{align}\label{final_control}
    \mathbf{u}=(\mathcal{B}^\top\mathcal{B})^{-1}\mathcal{B}^\top\left(\mathbf{B}{\mathbf{U}}_{\infty}-\mathbf{B}\mathbf{R}^{-1}\widetilde{\mathbf{B}}^{\top}\mathbf{P}\tilde{\mathbf{z}}-\mathbf{F}(\mathbf{x})\right)
\end{align}

Note that ${\mathbf{U}}_{\infty}$ and ${\mathbf{z}}_{\infty}$ are calculated through Eq. (\ref{steady_state}), $\mathbf{z}$ in $\tilde{\mathbf{z}}$ can be substituted by the designed measurement function $\mathbf{z}=\mathbf{g}(\mathbf{x},\mathbf{u})$ and $\mathbf{z}_{\rm I}$ can be directly obtained via the integrator (\ref{integrator}). Thus, the controller (\ref{final_control}) only requires feedback of $\mathbf{x}$ and is ready to be implemented in the original MG system (\ref{system_o}). The overall closed-loop MG control system based on the KO and LQI is shown in Fig. \ref{overall_control}.

\section{Case Studies}\label{Section:5}
This section presents several case studies that demonstrate the effectiveness of using the developed KO linearized model with the traditional LQI control method to stabilize the original nonlinear MG system and eliminate the steady-state error of DER output voltages caused by the droop equations.
	\subsection{ Simulation Setup}\label{simu_set}
The test system is a widely used $220$ V MG with three inverter-based DERs as shown in Fig. \ref{testfig} \cite{Pogaku2007}. The network is resistance-dominated for such a low-voltage distribution system. Table \ref{table_parameter} provides the parameter setting and initial states in this section. All three DERs are rated at 10 kVA with the same droop gain, so the load consumption is shared equally. Before the designed controller $\mathbf{u}$ in (\ref{final_control}) is applied, the voltage setpoints $v_{{\rm set}i}$ ($i=1,\dots,3$) in the droop equation (\ref{droop2}) for each DER are set as $380$ V, resulting in steady-state errors in DER output voltages ${v}_{{\rm od}i}$. All the dynamic simulations are conducted in the MATLAB environment. 
	
\begin{table}[t!]
\setlength{\tabcolsep}{5pt}
\centering
\caption{Parameter setting of MG}
\footnotesize
\renewcommand\arraystretch{1.5}
\label{table_parameter}
\begin{tabular}{p{0.6cm}<{\centering}|p{1cm}<{\raggedright}p{2.3cm}<{\raggedright}p{1cm}<{\raggedright}p{2.3cm}<{\raggedright}}
\hline\hline
& Par. & Value  & Par. & Value \\ \hline
& $\mathbf{v}_{\rm od}(0)$ & $[380.8, 381.8,380.4]$ & $\mathbf{v}_{\rm oq}(0)$ & $[0,0,0]$\\
 &  $\mathbf{i}_{\rm od}(0)$ & $[11.4,11.4,11.4]$ & $\mathbf{i}_{\rm oq}(0)$ & $[0.4,-1.45,1.25]$\\
Initial &  $\mathbf{i}_{\rm ld}(0)$ & $[11.4,11.4,11.4]$ & $\mathbf{i}_{\rm lq}(0)$ & $[-5.5,-7.3,-4.6]$\\ 
			& $\omega(0)$ & 314 &$\bm{\delta}_0$ & $[0,0.0019,-0.0113]$\\
			& $i_{{\rm line}1{\rm d}}(0)$ & -3.8 &$i_{{\rm line}1{\rm q}}(0)$ & 0.4\\
			& $i_{{\rm line}2{\rm d}}(0)$ & 7.6 &$i_{{\rm line}2{\rm q}}(0)$ & -1.3\\\hline
Line          &  $r_{\rm line1}$ & $0.23 \;\Omega $ & $x_{\rm line1}$ & $0.1 \;\Omega $\\
and   &  $r_{\rm line2}$ & $0.35 \;\Omega $ & $x_{\rm line2}$ & $0.58 \;\Omega $\\
Load		&  $r_{\rm load1}$ & $25 \;\Omega $ & $x_{\rm load3}$ & $20 \;\Omega $\\ \hline 
DER		&  \multicolumn{4}{l}{ The DER parameters can be found in \cite{Pogaku2007}}\\ \hline \hline
		\end{tabular}
	\end{table}
\subsection{Control Performance based on the KO and LQI}
		\begin{figure}[t!]
		\centering
		\includegraphics[width=0.7\columnwidth]{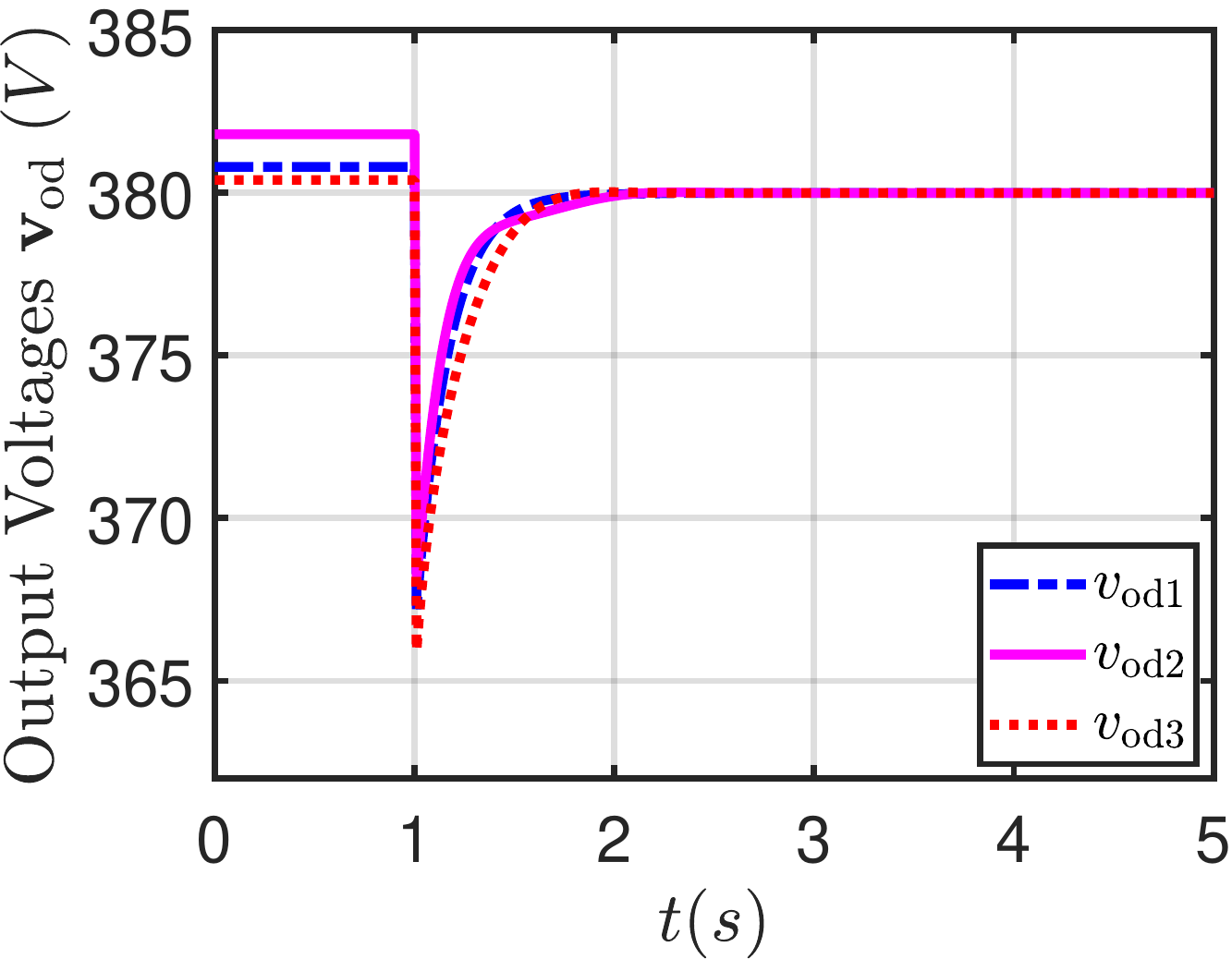}
		\caption{Dynamic responses of DER output voltages of the test MG.}
		\label{controlperformance}
						\vspace{-1 em}
	\end{figure}
The proposed KO linearized MG model for the voltage control of MGs is verified by applying the LQI controller (\ref{final_control}) to the original nonlinear MG model (\ref{system_o}) after $1$ s. Before that, the voltage setpoints for the droop equations are kept constant at $\mathbf{u}=[380, 380, 380]^\top$ V. Figure \ref{controlperformance} shows that the DER output voltages have steady-state errors due to the droop characteristic before $1$ s. When the proposed KO-based LQI controller takes over, the steady-state errors are quickly eliminated, confirming the effectiveness of the proposed method.

  		\begin{figure*}[t!]
		\centering
		\includegraphics[width=1.9\columnwidth]{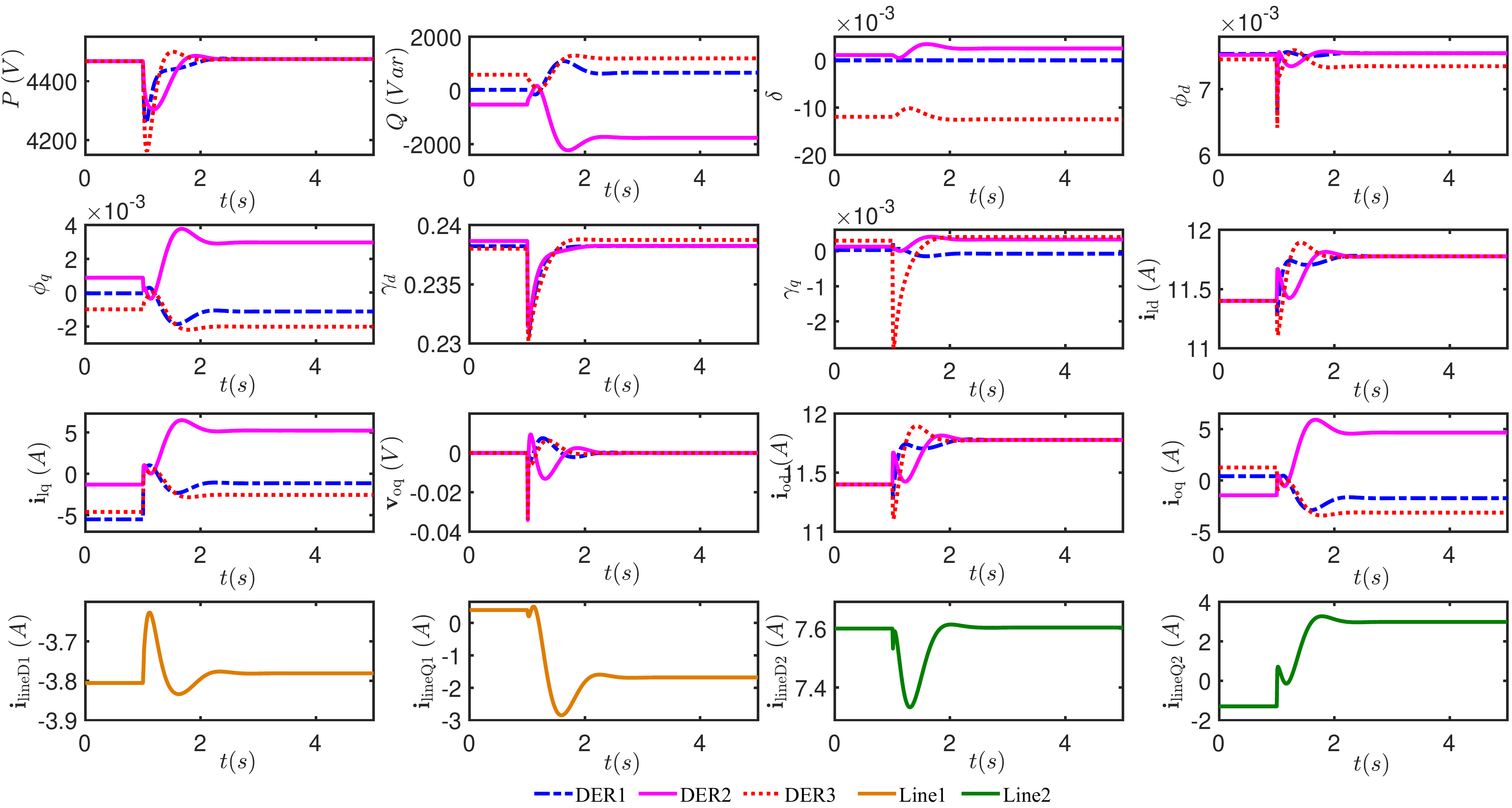}
		\caption{Dynamic responses of all the other state variables of the test MG.}
		\label{allstates}
						\vspace{-1 em}
	\end{figure*}
\begin{figure}[t!]
		\centering
		\includegraphics[width=0.9\columnwidth]{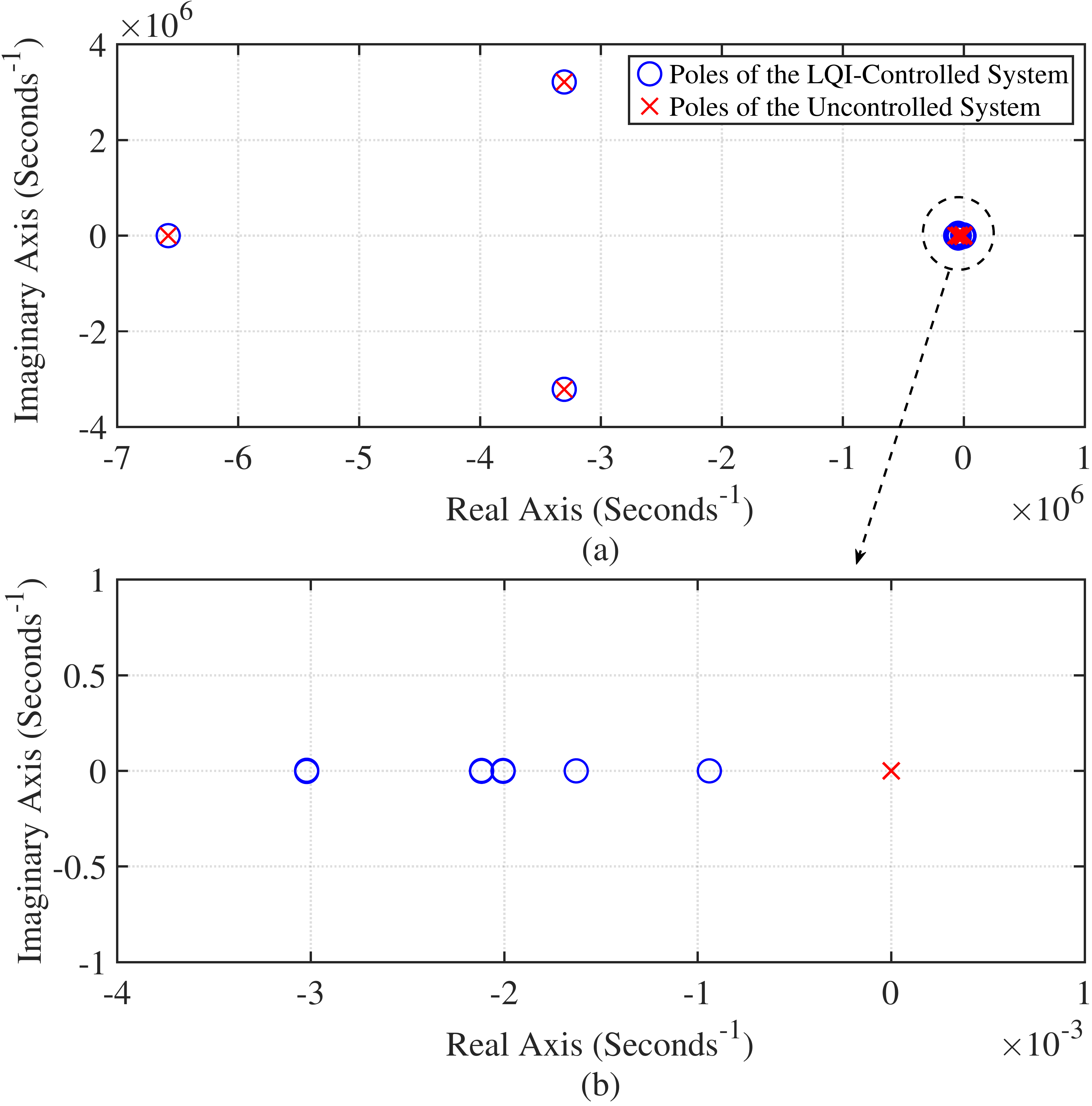}
		\caption{Comparison of poles of system (\ref{KO model}) before and after the LQI controller $\widetilde{\mathbf{U}}$ is applied.}
		\label{poles}
						\vspace{-1 em}
	\end{figure}

 Figure \ref{allstates} shows the dynamic responses of all the other stable variables. It can be observed that all the state variables are stabilized to a new equilibrium point. For a more systematic study of the system stability, we compare the poles of the system (\ref{KO model}) before and after the LQI controller $\widetilde{\mathbf{U}}$ are applied, i.e., eigenvalues of $\mathbf{A}$ and $\widetilde{\mathbf{A}}-\widetilde{\mathbf{B}}\mathbf{K}$. The maximum of the real part of eigenvalues of matrix $\mathbf{A}$ is $7.7709\times10^{-11}$ while that of matrix $\widetilde{\mathbf{A}}-\widetilde{\mathbf{B}}\mathbf{K}$ is $-9.4000\times10^{-4}$. However, it should be mentioned that the original nonlinear system (\ref{system_o}) is actually stable with the provided configuration. The reason that the KO linearized model (\ref{KO model}) has positive poles (indicating unstable modes) is that part of system dynamics $\mathbf{F}(\mathbf{x})$ is absorbed into the term $\mathbf{B}\mathbf{U}$ as discussed in Remark 1. Therefore, the poles of $\mathbf{A}$ only reflect the open-loop stability of the KO linearized system (\ref{KO model}), but do not indicate the stability of the original nonlinear system (\ref{system_o}). With the application of LQI, all the poles are placed on the plane's left side, indicating that the LQI controller stabilizes the system (\ref{KO model}) as shown in Fig. \ref{poles}. 

   		\begin{figure}[t!]
      \vspace{0.6 em}
		\centering
		\includegraphics[width=0.9\columnwidth]{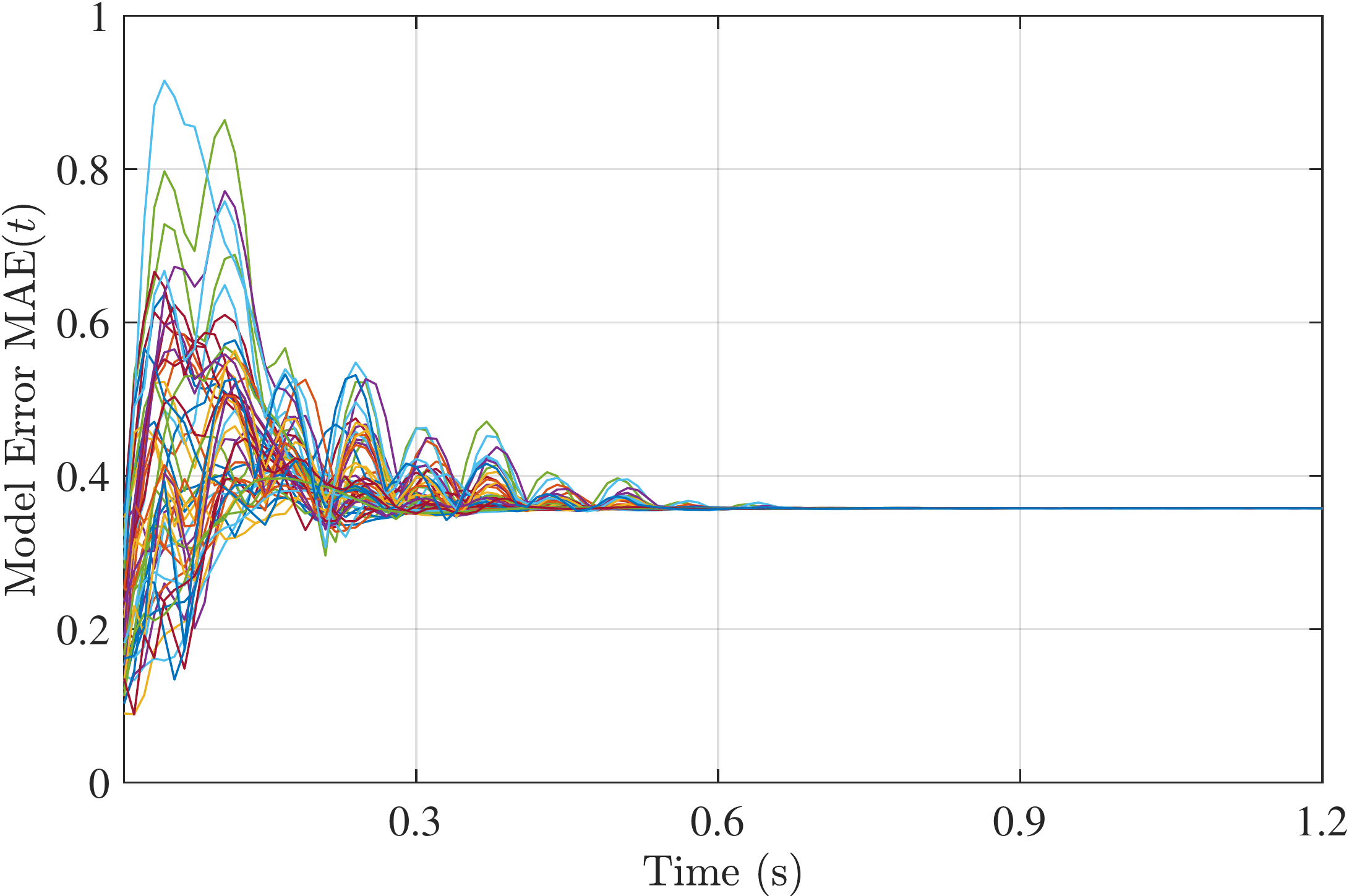}
		\caption{The time-varying model error measured by MAE with 50 different initial condition settings.}
		\label{modelerrors}
						\vspace{-1 em}
	\end{figure}
\subsection{Model Error and Sensitivity Analyses}\label{modelerror}
The KO linearized model (\ref{KO model}) is derived analytically, so the only source of model error between (\ref{KO model}) and (\ref{system_o}) should be the assumptions made in the model development, namely $\sin\delta_i\approx\delta_i$, $\cos\delta_i\approx1$, and $\omega_i\approx\omega_n$ in the LC filters and lines. To verify this claim, we set $\mathbf{u}=[380,380,380]^{\top}$ V for both (\ref{system_o}) and (\ref{KO model}). Since the observable vector $\mathbf{z}$ contains an explicit representation of the state vector of the original MG $\mathbf{x}$, we can denote the $\mathbf{x}$ in $\mathbf{z}$ as $\mathbf{z}_\mathbf{x}$. This allows us to directly compare the dynamic responses of the two models. Use mean absolute error (MAE) to define the model error as 
\begin{align}
    {\rm MAE}(t)=\frac{1}{n}\sum_{i=1}^{n}\left|\mathbf{x}(t)-\mathbf{z}_\mathbf{x}(t)\right|.
\end{align}

\begin{figure}[t!]
		\centering
		\includegraphics[width=0.9\columnwidth]{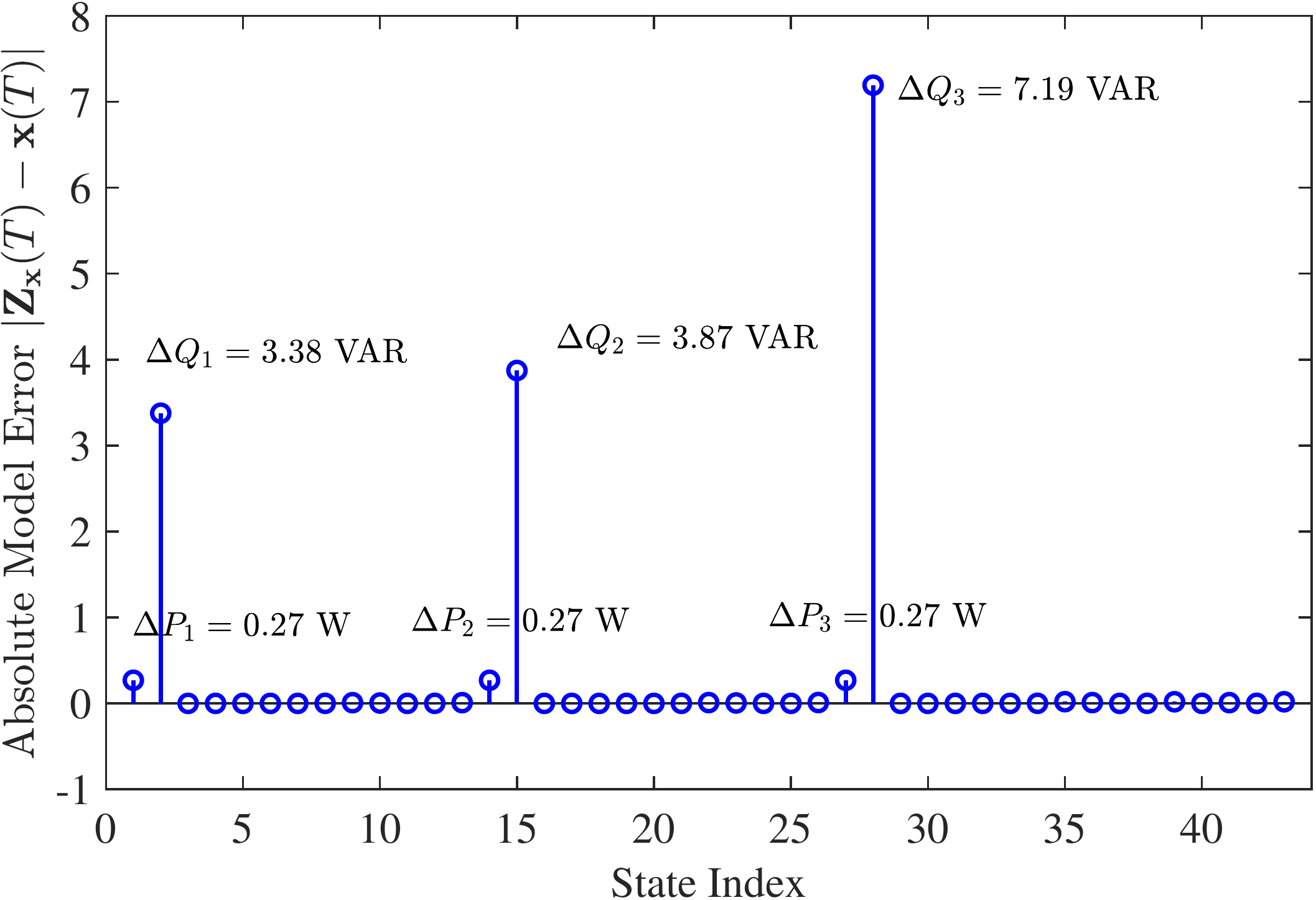}
		\caption{The steady-state absolute model error of each state at time $T=5$ seconds. $\Delta P$ and $\Delta Q$ denote the absolute error of real and reactive powers, respectively.}
		\label{stateerror}
	\end{figure}

We also conduct sensitivity analysis of the developed KO linearized model by simulating $50$ different sets of initial conditions. For each run, we add a $30\%$ random perturbation to the initial condition in Table \ref{table_parameter}. Figure \ref{modelerrors} shows that all the model errors MAE$(t)$ with different initial conditions oscillate during the settling period and finally converge to around $0.357$. Moreover, the MAE$(t)$ is always below $1$ throughout the timeline. To investigate the source of the steady-state error, we examine the detailed error of each state. Figure \ref{stateerror} reveals that the steady-state errors mainly occur in the active and reactive powers, but their actual values are negligible compared to the magnitude of $P$ and $Q$. Therefore, we conclude that the developed KO linearized model is sufficiently accurate and robust against different initial conditions.

\section{Conclusions}\label{Section:6}

This paper presents a novel large-signal method to linearize microgrid (MG) models for controller design using the Koopman operator (KO) theory. The primary and zero control levels are modeled for electromagnetic transient (EMT) analysis, which increases system order and nonlinearity. To overcome these challenges, we have derived the observable functions and KO analytically, avoiding data dependence and improving explainability. Voltage control with linear quadratic integrator (LQI) is used as an example to show how our KO linearized model enables textbook linear control techniques for nonlinear MGs. To guarantee stabilizability, a lifted-dimensional control signal has been derived in the KO linearized model. We use least squares to map the high-dimensional control vector to the original one. The case studies validate the LQI and KO linearized model for DER output voltage restoration. The model error without a state-feedback controller under different initial conditions confirms the accuracy and robustness of our analytical KO linearized MG model. The proposed analytical derivation methodology is generic and applicable to other MG systems with different structures and objectives.

\bibliographystyle{IEEEtran}
	\bibliography{reference}

\end{document}